# High-Throughput Study of Antisolvents on the Stability of Multicomponent Metal Halide Perovskites through Robotics-Based Synthesis and Machine Learning Approaches

Kate Higgins[1], Maxim. Ziatdinov[2,3], Sergei V. Kalinin[2] and Mahshid Ahmadi[1,ζ]

[1] *Joint Institute for Advanced Materials, Department of Materials Science and Engineering, The University of Tennessee Knoxville, Knoxville, TN 37996*

[2] *Center for Nanophase Materials Sciences, Oak Ridge National Laboratory, Oak Ridge, TN 37831*

[3] *Computational Sciences and Engineering Division, Oak Ridge National Laboratory, Oak Ridge, Tennessee 37831, USA*

ζ Corresponding author email: mahmadi3@utk.edu

**Abstract**

Antisolvent crystallization methods are frequently used to fabricate high-quality perovskite thin films, to produce sizable single crystals, and to synthesize nanoparticles at room temperature. However, a systematic exploration of the effect of specific antisolvents on the intrinsic stability of multicomponent metal halide perovskites has yet to be demonstrated. Here, we develop a high-throughput experimental workflow that incorporates chemical robotic synthesis, automated characterization, and machine learning techniques to explore how the choice of antisolvent affects the intrinsic stability of binary perovskite systems in ambient conditions over time. Different combinations of the endmembers, $MAPbI_3$, $MAPbBr_3$, $FAPbI_3$, $FAPbBr_3$, $CsPbI_3$, and $CsPbBr_3$, are used to synthesize 15 combinatorial libraries, each with 96 unique combinations. In total, roughly 1100 different compositions are synthesized. Each library is fabricated twice using two different antisolvents: toluene and chloroform. Once synthesized, photoluminescence spectroscopy is automatically performed every 5 minutes for approximately 6 hours. Non-negative Matrix Factorization (NMF) is then utilized to map the time- and compositional-dependent optoelectronic properties. Through the utilization of this workflow for each library, we demonstrate that the selection of antisolvent is critical to the stability of metal halide perovskites in ambient conditions. We explore possible dynamical processes, such as halide segregation, responsible for either the stability or eventual degradation as caused by the choice of antisolvent. Overall, this high-throughput study demonstrates the vital role that antisolvents play in the synthesis of high-quality multicomponent metal halide perovskite systems.

**Introduction**

Considerable research attention has focused on metal halide perovskites (MHPs) over the recent years because of the combination of exceptional optoelectronic properties and low fabrication cost, making them ideal candidates for a variety of applications, such as solar cells[1], photodetectors[2], ionizing radiation sensors[3, 4], and light-emitting diodes[5, 6]. Even so, the development of MHPs for commercialization must overcome an obstacle, namely stability in the pure or device-integrated form[7, 8]. Overcoming adverse effects stemming from external stimuli can be minimized or avoided by utilizing established encapsulation techniques and device engineering [9-11]. Simultaneously, another strategy is to improve the intrinsic stability by cation and/or halide alloying to synthesis multicomponent MHPs [12-14]. A multitude of studies have demonstrated how the incorporation of other cations, particularly cesium $(Cs^+)$[15] and formamidinium (FA+)[16] into methylammonium (MA) systems, leads to improve stability in ambient and operational conditions. Mixing halides, such as bromine $(Br^-)$[17], iodide $(I^-)$, and chlorine $(Cl^-)$[13], has also proven to be an effective strategy toward stable perovskite materials.

Antisolvent crystallization is an efficient solution-based method to produce high-quality perovskite thin films, to synthesize high-quality single crystals[18], and to fabricate nanoparticles at room temperature[19]. Simply, this approach requires the rapid application of the antisolvent to a perovskite solution, and the antisolvent then extracts the solvent, leading to the fast supersaturation of the perovskite precursor and subsequent precipitation. Many studies have focused on the investigation of antisolvents, including the types, volumes and the mixture of solvents[20]. For example, Sakai et al. demonstrated that dropping toluene on the perovskite precursor deposited film during spin-coating can enhance the early stages of nucleation and grain growth, depending on the chemical composition of the precursor solution[21]. Wang et al. postulated that the use of chlorobenzene mixed with iso-propyl alcohol removes residual chlorobenzene, therefore, increasing the grain size of the perovskite film and drop in defect density[22].

To date, much of these investigations have utilized manual trial and error approaches to decide which antisolvent is applicable for a particular perovskite system. Recently, automated experimentation has proven to accelerate this trial and error process through synthesis[23-25] and full device preparation workflows[26]. We have also previously reported the development of a workflow for materials discovery utilizing both automated synthesis and machine learning[27]. However, the number of groups utilizing these workflows to explore antisolvent engineering is small. Gu et al. utilized a robot-based high-throughput workflow to screen antisolvents on single component MHPs[24]. By testing 48 different organic ligands as an antisolvent and three commonly used solvents: dimethyl sulfoxide (DMSO), γ-butyrolactone (GBL), and *N'N*-dimethylformamide (DMF), they explore how the choice of solvent and antisolvent effect on the formation of microcrystals. As shown in another high-throughput study screening antisolvents by Manion et al., the quality of thin films can be predicted by analyzing these formations[28]. While both high-throughput studies focus on the effect of antisolvents on single endmember systems, studies exploring the effect of solvent engineering on multicomponent MHPs systems is limited, and none have explored the intrinsic stability of these compositions in ambient conditions.

Here, we utilize our previously reported workflow[27] to explore how the choice of antisolvent affects the photoluminescent (PL) behavior, and therefore stability, of multicomponent MHPs over time in ambient conditions. With the use of a pipetting robot, we synthesize 15 different binary perovskite systems, also referred to as a combinatorial library, twice utilizing two different solvents: toluene and chloroform. Overall, roughly 1100 unique perovskite compositions were synthesized. Next, PL spectroscopy is performed on each combinatorial library in ambient conditions for approximately 6 hours. Finally, through the incorporation of multivariate statistical analysis, specifically Non-negative Matrix Factorization (NMF), we map the time- and compositional-dependent PL behavior of each combinatorial library. Through the usage of this workflow for each combinatorial library, we characterize the intrinsic stability for each system in a specific antisolvent, helping guide future experiments done elsewhere.

**Results and Discussion**

We utilize an established workflow[27] by our group for the rapid synthesis and characterization of MHPs that combines combinatorial synthesis and rapid throughput photoluminescent (PL) measurements. Further, we adopt a multivariate statistical analysis explore the variability of the optical band gap and PL properties over time across the compositional space. A schematic of the workflow used is shown in **Figure 1a)**.

In this work, we explore 15 different multicomponent MHPs binary systems as shown in **Figure 1b)**. Our chosen endmembers are $MAPbI_3$, $MAPbBr_3$, $FAPbI_3$, $FAPbBr_3$, $CsPbI_3$, and $CsPbBr_3$. Synthesis details of the endmember solutions are provided in the **Supporting**

**Information**. We chose common solvents for each endmember, such as DMSO, GBL, and DMF to dissolve each precursor. In each combinatorial library, there are 96 unique compositions, meaning that roughly 1100 unique compositions were synthesized. We then utilize an antisolvent approach to precipitate microcrystals[4, 29]. Commonly used antisolvents documented by other works are classified into two categories: halogenated (chlorobenzene[30] and chloroform[29]) and non-halogenated (toluene[17], anisole[31], diethyl ether[32], and ethyl acetate[33]). Here, we chose to utilize two antisolvents: chloroform and toluene, as they represent each category and because these antisolvents are heavily used in the production of high-quality perovskites. Because we use two antisolvents, each combinatorial library is synthesized twice, and, therefore, in this study, roughly 2880 different microcrystal samples in total were synthesized.

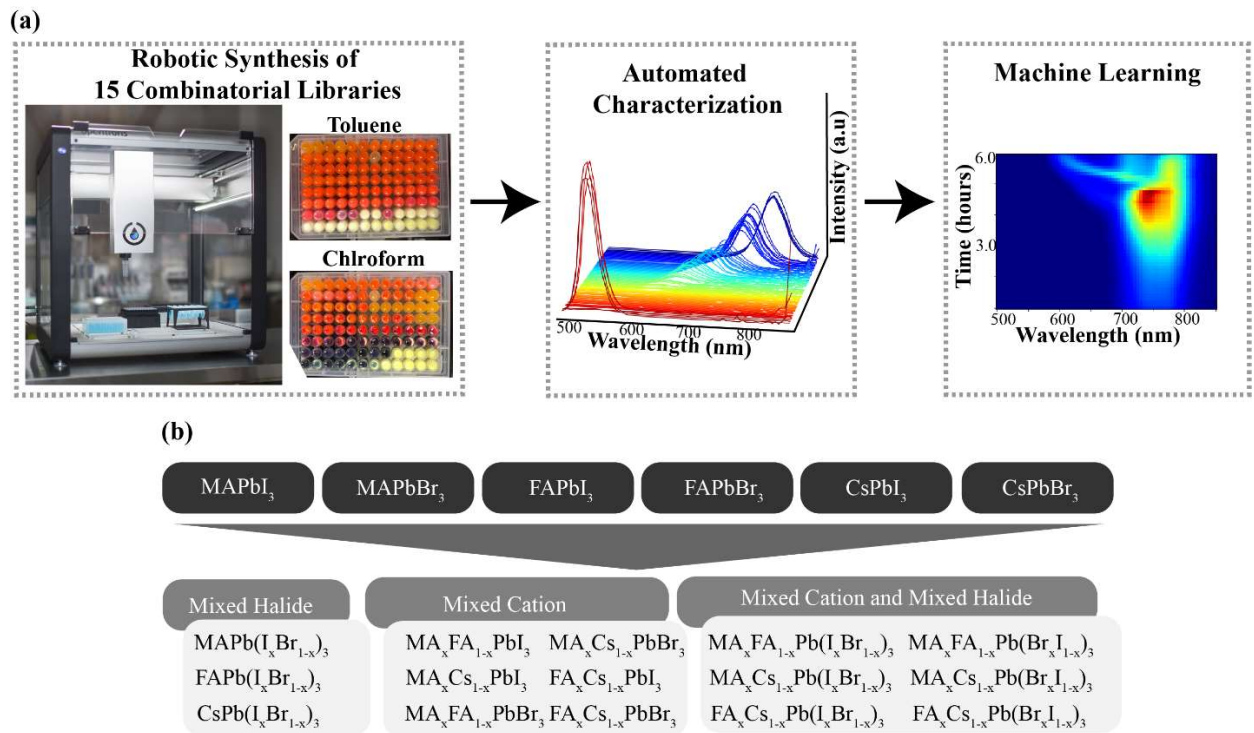

**Figure 1. a)** Schematic of the experimental workflow utilized for the exploration of 15 combinatorial libraries. We utilize robotic synthesis to fabricate a library twice with two different solvents: toluene and chloroform, photoluminescence (PL) spectroscopy is automatically performed next. Finally, we utilize machine learning to effectively map characteristic PL behaviors
5

onto composition regions. **b)** List of all combinatorial libraries that were chosen from combinations of the six endmembers.

First, utilizing a programmable pipetting robot (Opentrons: https://opentrons.com/) and employing classical 96-well microplates, we create the combinatorial libraries. We control the composition of each well by programming the robot to pipette the desired quantity of the endmember solution. After the deposition of the endmember solutions, the antisolvent is pipetted into each well. The formation of microcrystals is apparent by the color change that occurs immediately upon the addition of the antisolvent. Details of the synthesis and photos of selected microplates are provided in the **Supporting Information**.

To characterize compositional dependent optoelectronic properties, we utilize an automated Multi-Mode well plate reader with the capabilities to perform photoluminescence (PL). Details of these measurements are provided in the **Supporting Information**. By studying the evolution of optical properties, we extract information regarding dynamical processes related to MHPs formation, ionic movement, phase changes, and degradation[34-36] to characterize the stability of MHPs in ambient condition. While other common methods, such as X-ray diffraction (XRD)[36, 37], UV-vis absorption spectroscopy[34, 38], etc., can also be used to explore the intrinsic stability of MHPs, here, we utilize PL spectroscopy because it can be easily incorporated into automated workflows. We can measure the PL spectra of 96 compositions in roughly 5 minutes. We repeat this measurement for approximately 6 hours to obtain a time- and composition-dependent dataset for each combinatorial library.

To effectively map the time- and compositional-dependent PL properties and stability relationships concurrently, we adopt a multivariate statistical approach, in particular, Non-negative Matrix Factorization (NMF)[27]. Here, the entire dataset, $L(x, y, \omega, t)$, is analyzed via the decomposition,

$$L(x, y, \omega, t) = \sum_{i=1}^{N} L_i(x, y) g_i(\omega, t)$$

where $L_i(x,y)$ are the loading maps, representing the variability in PL behavior across the compositional space, and $g_i(\omega,t)$ are the endmembers that determine characteristic behaviors in the time-spectral domain. Here, the loading maps are 1D plot in which the *x*-axis represents compositional change and *y*-axis is the intensity, indicating if the endmember applies to the composition, i.e., low intensity, endmember less applicable. Endmember maps are 2D plots, demonstrating how the characteristic PL behavior changes over time.

As mentioned previously, the effect of an individual antisolvent, either toluene or chloroform, for crystallization was explored by characterizing a total of 15 combinatorial libraries each containing 96 unique compositions. For each plate, we have a time- and compositional-dependent dataset by automatically measuring the PL over time. While all combinatorial libraries are fully explored in the **Supporting Information**, here, we focus on only 4 combinatorial libraries: a mixed cation system, $MA_xFA_{1-x}PbI_3$, a mixed halide system, $MAPb(I_xBr_{1-x})_3$, and two mixed cation and mixed halide systems, $MA_xCs_{1-x}Pb(I_xBr_{1-x})_3$ and $MA_xCs_{1-x}Pb(Br_xI_{1-x})_3$.

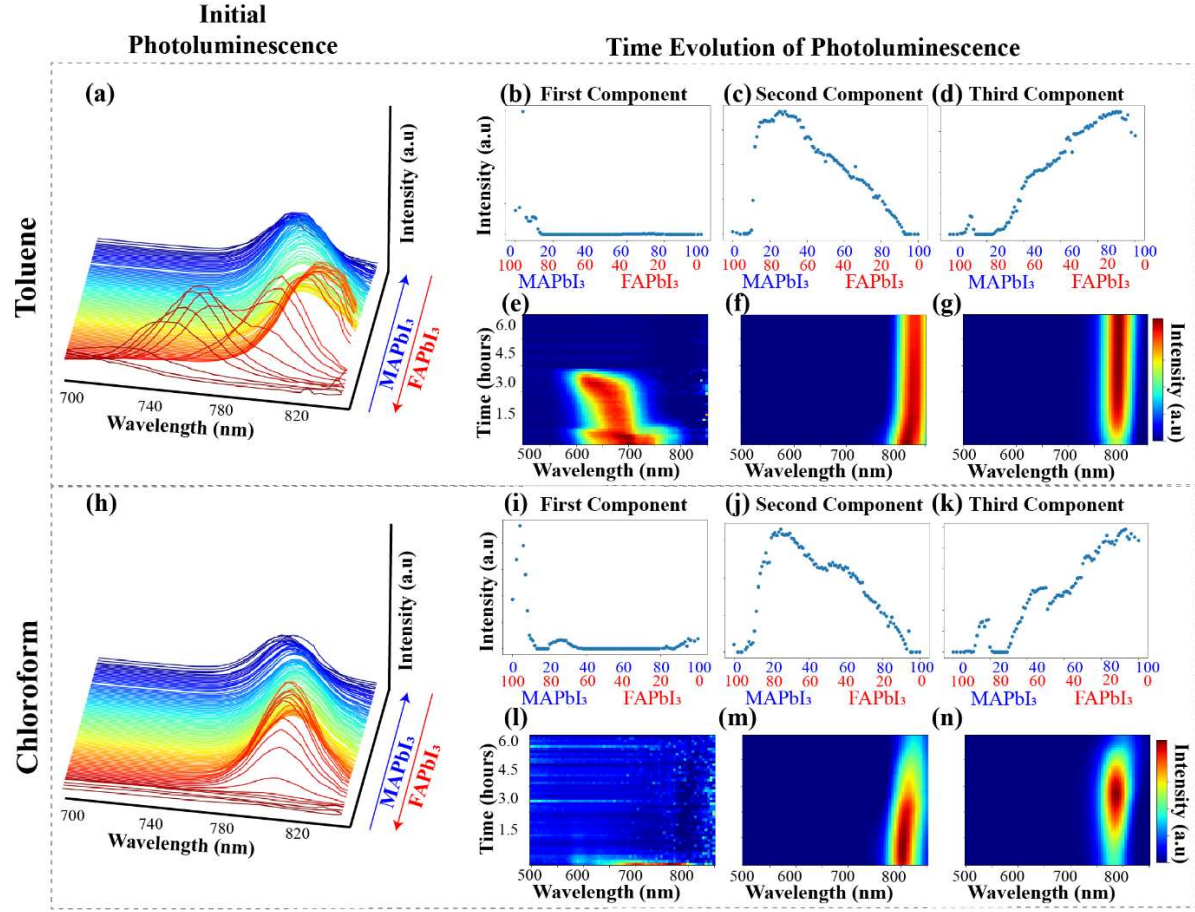

**Figure 2. Double cation lead iodide system, $MA_xFA_{1-x}PbI_3$. a)** initial PL behavior when toluene is used as the antisolvent. **b), c)**, and **d)** are the loading maps for $FAPbI_3$-rich compositions, solid solutions of $FAPbI_3$ and $MAPbI_3$, and $MAPbI_3$-rich compositions, respectively. **e), f)**, and **g)** Characteristic PL behavior for $FAPbI_3$-rich compositions, solid solutions of $FAPbI_3$ and $MAPbI_3$, and $MAPbI_3$-rich compositions, respectively. **h)** initial PL behavior when chloroform is used as the antisolvent. Similar to NMF decomposition for the toluene data set **i), j)**, and **k)** and **l), m)**, and **o)** are the loading maps and characteristic PL behavior for $FAPbI_3$-rich compositions, solid solutions of $FAPbI_3$ and $MAPbI_3$, and $MAPbI_3$-rich compositions, respectively.

Firstly, we begin with the exploration of the initial PL behavior of the mixed cation system, $MA_xFA_{1-x}PbI_3$, as shown in **Figures 2a)** and **2h)**. Regardless of the antisolvent used, we observe the shifting of the PL peak as the dopants increases, confirming compositional tuning of the bandgap[39]. Noticeably, as shown in **Figure 2a)**, the $FAPbI_3$-rich compositions have a broad photoluminescence peak and is shifted towards lower wavelengths. Based on this PL behavior and the yellow color of the microcrystals upon application of the antisolvent (**Figure S2**), we assume that the photoinactive, yellow phase ($\delta$-FAPbI$_3$) has formed in combination with the black phase ($\alpha$-FAPbI$_3$). Further, as shown in **Figure 2a)**, the formation of this yellow phase begins to recede when the concentration of MAPbI$_3$ increases, confirming stabilization upon doping[40]. We also observe this phenomenon when chloroform is utilized as the antisolvent; however, as the peak intensities are low compared to the other spectra, it is not evident in **Figure 2h)**. Shown in **Figure SX** are the spectrum of these compositions to confirm. From this initial measurement, we reach two conclusions: 1) neither toluene nor chloroform is the appropriate antisolvent choice for the formation of a pure phase $\alpha$-FAPbI$_3$, and 2) stabilization of the $\alpha$-FAPbI$_3$ phases can be achieved through small amount of MAPbI$_3$ in agreement with previous studies[40].

Further, by exploring the time-dependent PL behavior, we conclude this double cation system is vastly more stable when toluene is utilized as the antisolvent instead of chloroform. As mentioned previously, the FAPbI$_3$-rich compositions are likely a solid solution between the $\delta$-FAPbI$_3$ and $\alpha$-FAPbI$_3$ phases; therefore, when we examine the characteristic PL behaviors (**Figures 2e)** and **2l)** and corresponding loading maps, **Figures 2b)** and **2i)**) when either toluene or chloroform is used, we observe a broad peak at lower wavelengths spanning approximately 100 nm. With chloroform, this peak disappears quickly after roughly 30 minutes, whereas with the toluene system, we see this peak begin to narrow and shift towards lower wavelengths, indicating better stability of these mixed phases in toluene rather than chloroform. Next, we observe a similar trend when comparing the solid solution between FAPbI$_3$-rich and MAPbI$_3$ compositions in toluene and chloroform. Even with a small peak shift and slight decrease in intensity, toluene as the antisolvent produces a stable PL peak for the entirety of the characterization (roughly 6 hours in ambient conditions), as shown in **Figure 2f)**. Other studies have confirmed that the stabilization of $\alpha$-FAPbI$_3$ in ambient conditions can be achieved through small amounts of doping with MAPbI$_3$[40]. Conversely, this stabilization is temporary when chloroform is used as the antisolvent, as shown in **Figure 2m)**. For approximately 4 hours, we observe a steady PL peak before it begins to vanish. We attribute this decrease in PL intensity to the perovskite reacting with the moisture in ambient conditions, possibly causing the sample to revert to PbI$_2$ or forming a hydrate product[41]. Our conclusions that toluene is a better choice for this system is further confirmed by analyzing the time-dependent PL behavior of the MAPbI$_3$-rich compositions. As shown in **Figures 2d)** and **2g)**, these compositions demonstrate remarkable stability as their peak properties, such as position and width, remain constant throughout the characterization even after a small increase in PL intensity at the beginning, indicating possible filling of trap states[42]. A similar increase in PL intensity is observed for the system when chloroform is used, as shown in **Figures 2k)** and **2n)**; however, after approximately 5 hours, the PL intensity decreases, indicating the formation of hydrate products is occurring for these samples as well[41]. Overall, we conclude this system, except for the FAPbI$_3$-rich compositions that formed two phases, are remarkably stable with the utilization of toluene as the antisolvent. We postulate that chloroform does not fully extract the GBL, and, therefore, leaves unreacted precursors to interact with the air, eventually causing degradation.

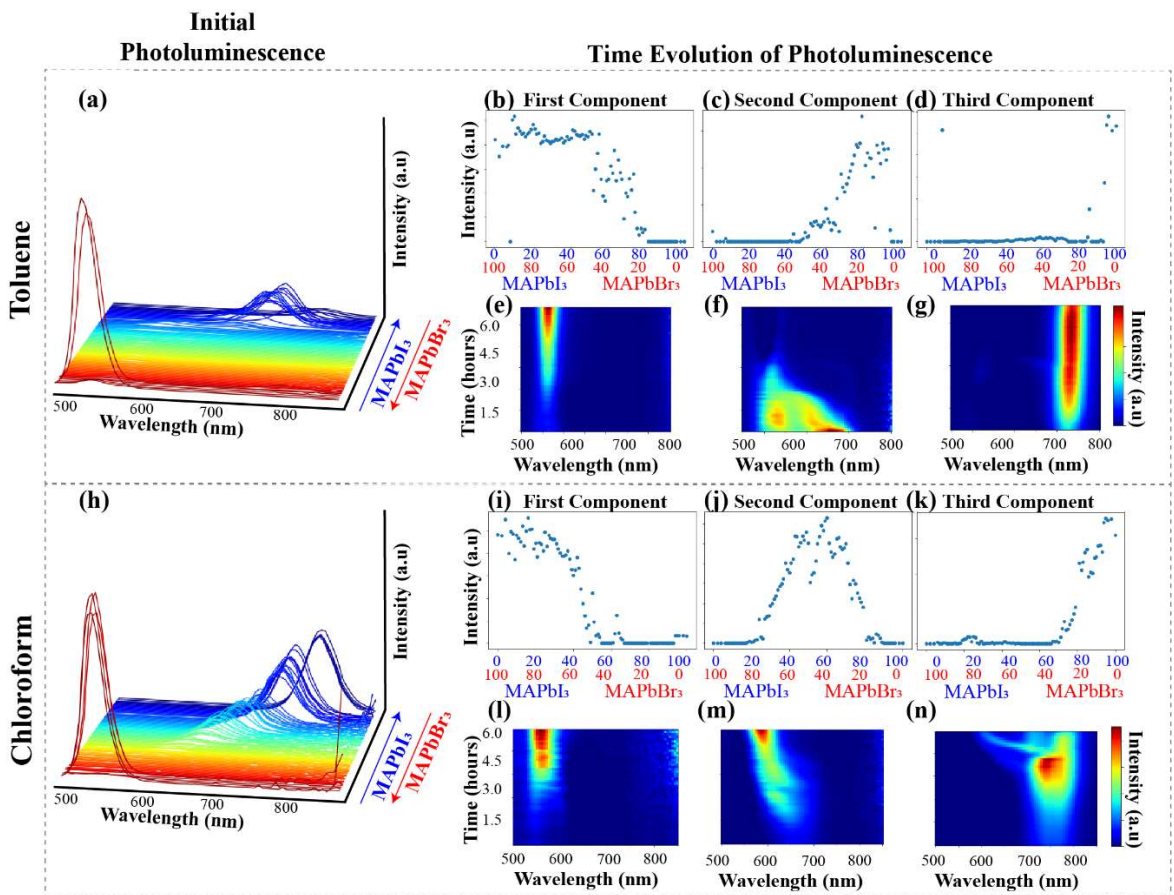

**Figure 3**. **Methylammonium lead double halide system, MAPb(I$_x$Br$_{1-x}$)$_3$. a)** initial PL behavior when toluene is used as the antisolvent. **b)**, **c)**, and **d)** are the loading maps for MAPbBr$_3$-rich compositions, solid solutions of MAPbBr$_3$ and MAPI$_3$, and MAPbI$_3$-rich compositions,

respectively. **e)**, **f)**, and **g)** Characteristic PL behavior for MAPbBr$_3$-rich compositions, solution solutions of MAPbBr$_3$ and MAPbI$_3$, and MAPbI$_3$-rich compositions, respectively. **h)** initial PL behavior when chloroform is used as the antisolvent. Similar to NMF decomposition for the toluene data set **i)**, **j)**, and **k)** and **l)**, **m)**, and **o)** are the loading maps and characteristic PL behavior for MAPbBr$_3$-rich compositions, solution solutions of MAPbBr$_3$ and MAPbI$_3$, and MAPbI$_3$-rich compositions, respectively.

Next, we further proceed by investigating how the change in antisolvent affects a mixed halide perovskite system: MAPb(I$_x$Br$_{1-x}$)$_3$. As shown in **Figures 3a)** and **3h)**, we observe a peak shift toward higher wavelengths as the concentration of MAPbI$_3$ increases, indicating an expected decrease in the bandgap. This shift is more apparent when chloroform is utilized as the antisolvent as the MAPbI$_3$-rich compositions have comparatively a large PL intensity as compared to the same compositions when toluene was used.

Regardless of the antisolvent, the PL behavior of MAPbBr$_3$-rich compositions is approximately the same; however, this is not applicable as more MAPbI$_3$ is introduced into the system. For MAPbBr$_3$-rich compositions, we observe an increase in PL intensity without any shift in peak position for both toluene (**Figures 3b)** and **3e)**) and chloroform (**Figures 3i)** and **3l)**), indicating that these compositions are stable regardless of antisolvent choice. Solid solutions between MAPbBr$_3$ and MAPbI$_3$ experience two different dynamical processes depending on the antisolvent used during synthesis. For example, as shown in **Figures 3c)** and **3f)**, a broad peak, likely indicative of two convolved peaks, is present initially. This peak begins to separate into two, one representing the bromine-rich phase and mixed phase. After approximately 3 hours, both peaks disappear, indicating complete degradation of the compositions. Simultaneously, when these same compositions are synthesized using chloroform as the antisolvent, we observe the presence of a broad, low-intensity peak. Over time, it appears to narrow as the intensity of the bromine-rich phase increases and the mixed-phase decreases. In both cases, we observe halide segregation caused by reactions with the environment[35], indicating this halide segregation cannot be prevented for these particular composition ranges. Finally, MAPbI$_3$-rich compositions in this system behave similarly to the same region of the phase diagram for the MA$_x$FA$_x$PbI$_3$, as expected. As shown in **Figures 2d)** and **2g)**, the MAPbI$_3$-rich compositions, specifically x ≤ 10%, experience an increase in PL intensity, but remain stable for the entirety of the experiment when toluene is utilized. As with the MAPbI₃-rich compositions when chloroform is used, we observe an increase and subsequent decrease in PL intensity as they are exposed to air, as shown in **Figures 2k)** and **2n)**. Interestingly, we observe a small peak extending towards lower wavelengths, indicating halide segregation occurring even with small amount of MAPbBr₃. Overall, this system is particularly prone to halide segregation, and neither antisolvent can improve this phenomenon. Either antisolvent produces a stable, well-defined PL peak for MAPbBr₃-rich compositions; however, toluene is the more appropriate choice when incorporating small amounts of MAPbBr₃ into MAPbI₃.

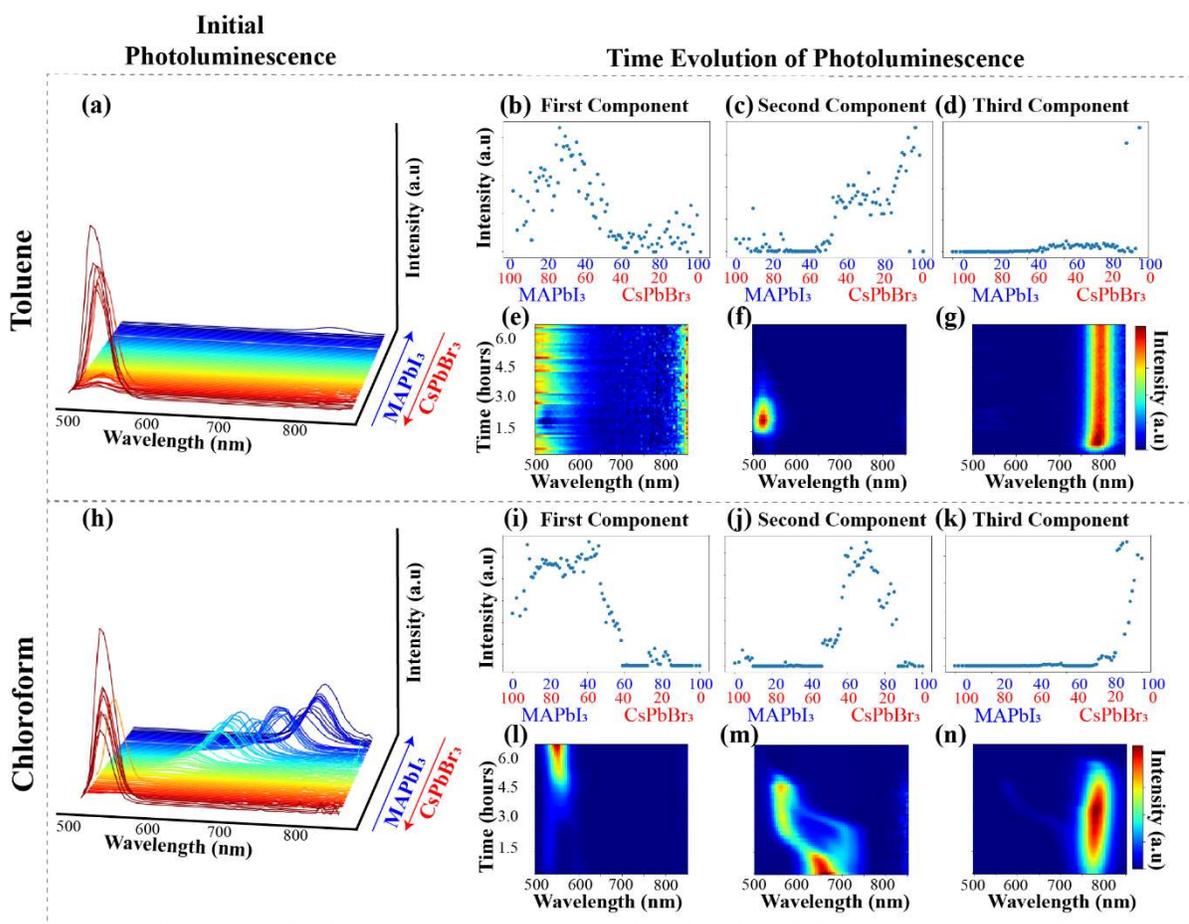

**Figure 4**. **Double cation and double halide system, MA$_x$Cs$_{1-x}$Pb(I$_x$Br$_{1-x}$)$_3$. a)** initial PL behavior when toluene is used as the antisolvent. **b)**, **c)**, and **d)** are the loading maps for CsPbBr₃-rich compositions, solution solutions of CsPbBr₃ and MAPbI₃, and MAPbI₃-rich compositions,

respectively. **e)**, **f)**, and **g)** Characteristic PL behavior for CsPbBr$_3$-rich compositions, solution solutions of CsPbBr$_3$ and MAPbI$_3$, and MAPbI$_3$-rich compositions, respectively. **h)** initial PL behavior when chloroform is used as the antisolvent. Similar to NMF decomposition for the toluene data set **i)**, **j)**, and **k)** and **l)**, **m)**, and **o)** are the loading maps and characteristic PL behavior for CsPbBr$_3$-rich compositions, solution solutions of CsPbBr$_3$ and MAPbI$_3$, and MAPbI$_3$-rich compositions, respectively.

We proceed further by investigating the effect of each antisolvent on the double cation and double halide system: MA$_x$Cs$_{1-x}$Pb(I$_x$Br$_{1-x}$)$_3$, as shown in **Figure 4.** As mentioned previously, we compare this system to a similar one, which uses different precursors, MA$_x$Cs$_{1-x}$Pb(Br$_x$I$_{1-x}$)$_3$, (**Figure 5**). Firstly, as shown in **Figures 4a)** and **4h)**, we clearly observe that MAPbI$_3$-rich compositions are comparatively lower in intensity to CsPbBr$_3$-rich compositions when toluene is used, whereas this is not the case when chloroform is utilized as we can see high intensity MAPbI$_3$ peaks. However, in both cases, adherence to Vegard's law, in which the bandgap shifts based on compositional doping, is upheld.

Regarding the time-dependent PL behavior, from the NMF decompositions, it becomes apparent that toluene is a more beneficial antisolvent for MAPbI$_3$ and chloroform is a better antisolvent for CsPbBr$_3$. As shown in **Figures 4b)** and **4e)**, rapid degradation of the CsPbBr$_3$-rich compositions is observed when toluene is used because the individual PL spectra could simply be characterized as noise, as shown in **Figure S3.** Conversely, when chloroform is utilized as the antisolvent, for these same compositions, we observe an increase in peak intensity without major peak shifts, indicating the stability of these compositions over time in ambient conditions. Next, we explore how each antisolvent effected the stability of solid solutions of MAPbI$_3$-rich and CsPbBr$_3$. For toluene, we observe a short appearance of an intense peak after approximately 2 hours before it disappears, as shown in **Figures 4c)** and **4f)**. Noticeably, even though these compositions contain a considerable amount of iodide, we do not observe an iodide-rich phase PL peak. On the contrary, when chloroform is used, we initially observe a mixed-phase PL peak, as shown in **Figure 4j)** and **4f)**; however, after approximately 2 hours, this peak separates into two peaks, each representing the iodide-rich and bromide-rich phases, confirming halide segregation[35]. Interestingly, the iodide-rich peak appears not to disappear as time increases but begins to shift towards lower wavelengths. Finally, as stated previously, MAPbI$_3$-rich compositions are remarkably stable with toluene as the antisolvent, as shown in **Figures 4d)** and **4g)**, whereas when chloroform is used, these same compositions experience an increase in PL intensity before another decrease, as shown in **Figures 4k)** and **4n)**. Overall, although it was not unexpected that for CsPbBr$_3$-rich compositions, toluene is not the appropriate antisolvent choice because of the polarity index disparity between DMSO and toluene[43], we were surprised by the finding that toluene was not the ideal antisolvent for solid solutions between CsPbBr$_3$ and MAPbI$_3$, considering there have been studies showing how toluene produces high-quality thin films when a combination of DMSO and GBL is used[17].

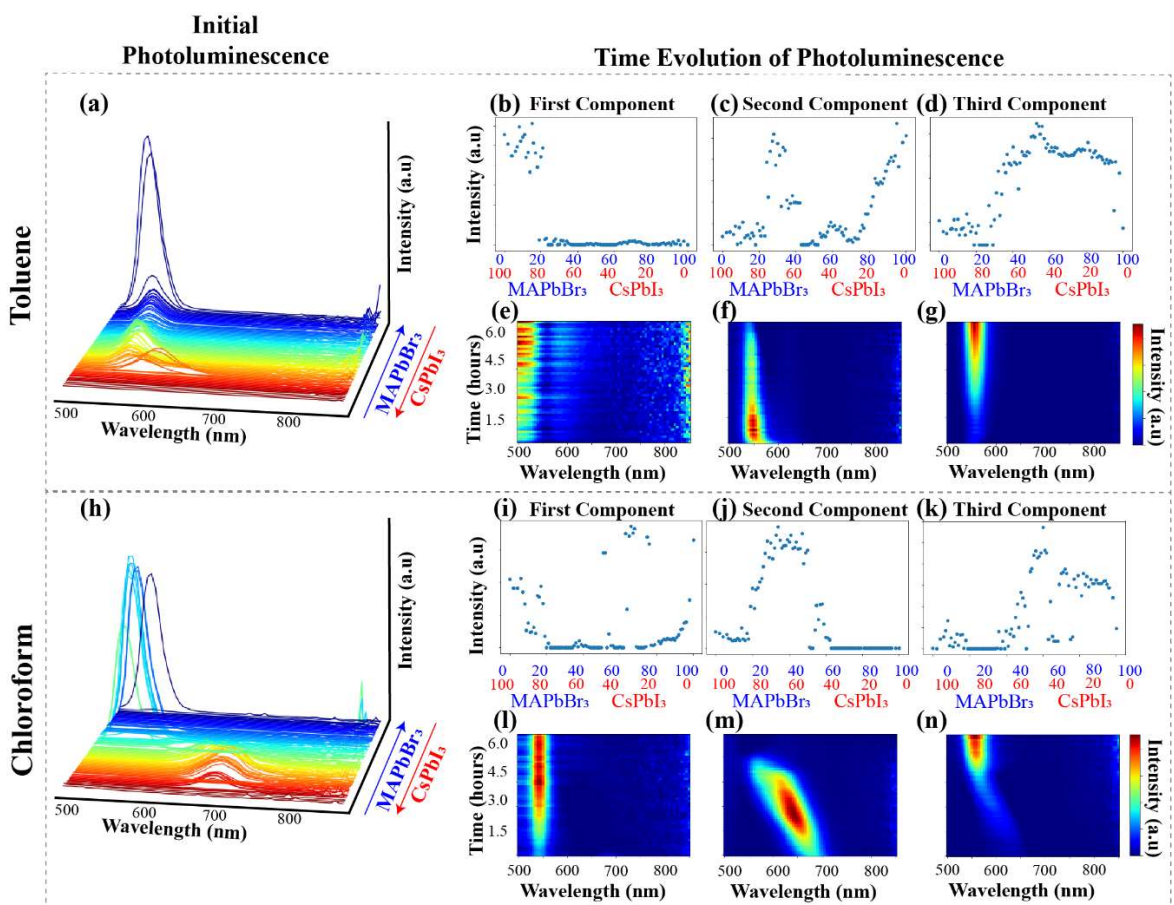

**Figure 5**. **Double cation and double halide system, MA$_x$Cs$_{1-x}$Pb(Br$_x$I$_{1-x}$)$_3$. a)** initial PL behavior when toluene is used as the antisolvent. **b)**, **c)**, and **d)** are the loading maps for CsPbI$_3$-rich compositions, solution solutions of CsPbI$_3$ and MAPbBr$_3$, and MAPbBr$_3$-rich compositions,

respectively. **e)**, **f)**, and **g)** Characteristic PL behavior for CsPbI$_3$-rich compositions, solution solutions of CsPbI$_3$ and MAPbBr$_3$, and MAPbBr$_3$-rich compositions, respectively. **h)** initial PL behavior when chloroform is used as the antisolvent. Similar to NMF decomposition for the toluene data set **i)**, **j)**, and **k)** and **l)**, **m)**, and **o)** are the loading maps and characteristic PL behavior for CsPbI$_3$-rich compositions, solution solutions of CsPbI$_3$ and MAPbBr$_3$, and MAPbBr$_3$-rich compositions, respectively.

Further, we explore a similar system as **Figure 4** by utilizing different precursors to analyze the PL behavior of the system: MA$_x$Cs$_{1-x}$Pb(Br$_x$I$_{1-x}$)$_3$. When either toluene or chloroform is utilized, the initial PL behavior for MAPbBr$_3$-rich compositions is a well-defined, high intensity peak, as shown in **Figures 5a)** and **5h)**. For CsPbI$_3$-rich compositions, there is a broad low-intensity peak observed when toluene is used, as shown in **Figure S4a)**, indicating this material not only formed the black phase of CsPbI$_3$, but also the wide-band gap, polymorph δ-CsPbI$_3$[44]. The formation of this phase is not dependent on the antisolvent, as evidence by its presence when chloroform is used, as shown in **Figure S4b)**, but instead, this phase is introduced by exposure to ambient conditions. Noticeably, Vegard's Law is better adhered to when chloroform is utilized versus the usage of toluene.

Regarding the time-dependent evolution of the PL for this system, we observe a variety of dynamical processes caused by either antisolvent. Firstly, the formation of CsPbI$_3$, most likely the polymorph phases, when toluene is used produces an instable composition, as evidence by **Figures 5b)** and **5e)**. Surprisingly, chloroform produces a stable δ-CsPbI$_3$, as shown in **Figures 5i)** and **5l)**. The band gap of this polymorph is roughly 2.82 eV[44], explaining the apparent peak at lower wavelengths, shown in **Figure 5l)**. Next, solid solutions between rich CsPbI$_3$-rich and MAPbBr$_3$ compositions exhibit a decreasing PL intensity at one peak wavelength over the course of time, indicating degradation of the compositions, as shown in **Figures 5c)** and **5f)**. Conversely, these same compositions begin with a single mixed-phase PL peak before experiencing halide segregation as the peak begins to move towards lower wavelengths, as shown in **Figures 5j)** and **5m)**. Finally, we observe similar trends, in which the peak goes in intensity of time, for the MAPbBr$_3$-rich compositions for both antisolvents (**Figures 5d)**, **5g)**, **5k), and 5n))**. However, as shown in **Figure 5k)**, when chloroform is used, these compositions also include far amounts of CsPbI$_3$, explaining the peak shift over time, as shown in **Figure 5n)**. Full stability analysis of this system is limited by the formation of δ-CsPbI$_3$ as it does not allow us to fully explore dynamical processes, such as halide segregation, when toluene is used; however, chloroform appears to be the appropriate choice to help with the formation of α-CsPbI$_3$ through the doping of MAPbBr$_3$.

Finally, we compare the solid solutions of the system, MA$_x$Cs$_{1-x}$Pb(I$_x$Br$_{1-x}$)$_3$, and the system, MA$_x$Cs$_{1-x}$Pb(Br$_x$I$_{1-x}$)$_3$, to compare the effect of the precursors as related to the antisolvent choice. To clarify, in the system, MA$_x$Cs$_{1-x}$Pb(I$_x$Br$_{1-x}$)$_3$, MAPbI$_3$ and CsPbBr$_3$ are the precursors, while in the system, MA$_x$Cs$_{1-x}$Pb(Br$_x$I$_{1-x}$)$_3$, the precursors are MAPbBr$_3$ and CsPbI$_3$. For both systems, it is demonstrated that toluene is not the appropriate choice for solid solutions of CsPbX$_3$-rich and methylammonium lead halide. In the MA$_x$Cs$_{1-x}$Pb(I$_x$Br$_{1-x}$)$_3$ system, the polar index disparity between DMSO and toluene causes the formation of CsPbBr$_3$ to be stagnated; whereas toluene produces an instable δ-CsPbI$_3$ phases in the MA$_x$Cs$_{1-x}$Pb(Br$_x$I$_{1-x}$)$_3$ system. Conversely, in both systems, chloroform produces initial mixed phase compositions; however, eventually, halide segregation begins to occur. Overall, this comparison is indicative of the precursor having an effect of the types of dynamical processes to occur as caused by the solvent in which the endmember is dissolved.

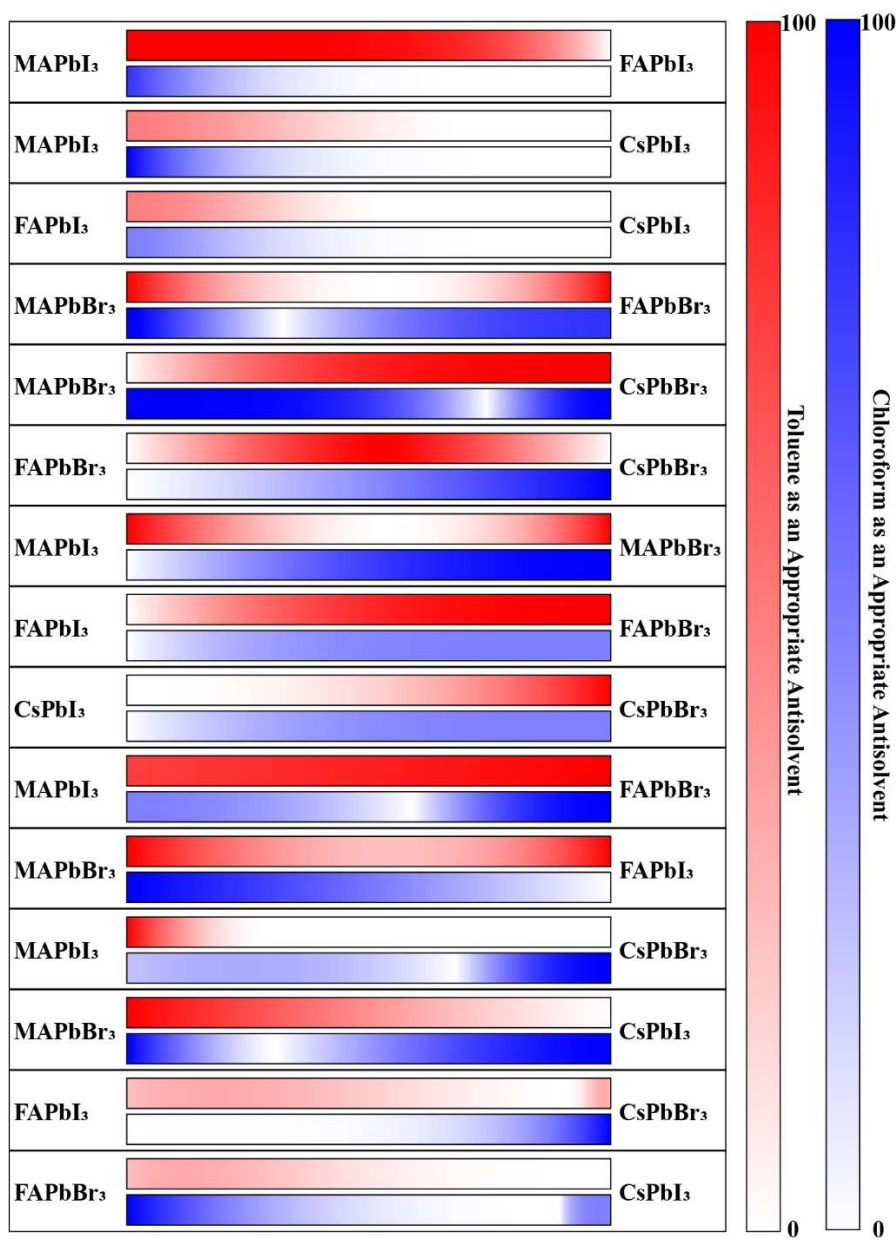

**Figure 6**. **Assessment for which antisolvent is more applicable for each system.** White or light red/blue indicate the antisolvent is not ideal for that section of the phase diagram as the microcrystals either form different phases or exhibit poor stability. As the color darkens, this indicates the antisolvent is an appropriate choice for the portion of the phase diagram as the correct phase forms and remains stable for longer periods of time.

In summary, we have demonstrated the effect of antisolvents on the initial photoluminescence behavior for multicomponent MHPs and on the dynamical processes experienced by these compositions over time when exposed to ambient conditions. Through the utilization of robotic synthesis, automated characterization, and machine learning, we effectively explore this effect on 15 different binary systems, analyzing approximately 1100 unique compositions. **Figure 6** provides a summary for these systems and indicates which antisolvent provides better stability in ambient conditions over time.

We have shown remarkable stability in ambient conditions for microcrystals of $MAPbI_3$-rich compositions and solid solutions between $FAPbI_3$ rich and $MAPbI_3$ compositions when toluene is utilized. We observed different phases of $FAPbI_3$ and $CsPbI_3$ and demonstrated how antisolvents influence the stability of those phases. Dynamical processes, such as halide segregation, were explored for many systems. Finally, we compared the same compositions synthesized with different precursor solutions and illustrated how the precursor can also play a role in the stability.

**Conclusion**

Through the utilization of our previously reported workflow[27], we have demonstrated how it can be used to synthesize and characterize large amounts of compositions to explore both the effect of experimental choices and dynamical process caused by exposure to ambient conditions. As stated previously, we synthesized and characterized approximately 2880 different microcrystal samples, and all together the absolute synthesis and characterization time total only 1.5 weeks. Through this procedure, we made conclusions about which antisolvent was the appropriate choice for a particular compositional region, providing guidance for future studies by exploring the intrinsic material stability.

**Acknowledgements**

M.A. acknowledges support from National Science Foundation (NSF), Award Number # 2043205. This research was partially supported by StART UTK-ORNL science alliance program. M.A., K.H. acknowledge support from CNMS user facility, project# CNMS2019-268. K.H. was partially supported by the Center for Materials Processing, a Center of Excellence at the University of Tennessee, Knoxville funded by the Tennessee Higher Education Commission (THEC). The GP process development (MZ, SKV) was supported by the Center for Nanophase Materials Sciences, which is a US DOE Office of Science User Facility.

# Supporting Information

# High-Throughput Study of Antisolvents on the Stability of Multicomponent Lead Halide Perovskites through Robotics-Based Synthesis and Machine Learning Approaches

Kate Higgins[1], Maxim. Ziatdinov[2,3], Sergei V. Kalinin[2] and Mahshid Ahmadi[1,ζ]

[1] *Joint Institute for Advanced Materials, Department of Materials Science and Engineering, The University of Tennessee Knoxville, Knoxville, TN 37996*

[2] *The Center for Nanophase Materials Sciences, Oak Ridge National Laboratory, Oak Ridge, TN 37831*

[3] *Computational Sciences and Engineering Division, Oak Ridge National Laboratory, Oak Ridge, Tennessee 37831, USA*

[ζ]**Corresponding author email: mahmadi3@utk.edu**

**Materials**

Methylammonium iodide (Sigma-Aldrich, ≥99%, anhydrous), methylammonium bromide (Sigma-Aldrich, ≥99%, anhydrous), formamidinium iodide (Sigma Aldrich, ≥99%, anhydrous), formamidinium bromide (Sigma-Aldrich, ≥99%, anhydrous), cesium iodide (Sigma-Aldrich, anhydrous, beads, −10 mesh, 99.999% trace metals basis), cesium bromide (Sigma-Aldrich, 99.999% trace metals basis), lead iodide (Sigma-Aldrich, 99.999% trace metals basis) and lead bromide (Sigma-Aldrich, 99.999% trace metals basis) were used without further purification.

**Methods**

*Precursor Solution Synthesis*

We made 0.3 M solutions by dissolving the precursor materials into the solvent shown in **Table S1**. The precursor solutions were left to stir for approximately one hour before they were removed and immediately used for the robotic synthesis.

| Material | Solvent |
|---|---|
| MAPbI$_3$ | γ-butyrolactone (GBL) |
| MAPbBr$_3$ | *n'n*-dimethylformamide (DMF) |
| FAPbI$_3$ | γ-butyrolactone (GBL) |
| FAPbBr3 | *n'n*-dimethylformamide (DMF) |
| CsPbI$_3$ | *n'n*-dimethylformamide (DMF) |
| CsPbBr$_3$ | Dimethyl sulfoxide (DMSO) |

**Table S1.** Chosen solvent for each material synthesized.

*Robotic Synthesis*

Automated synthesis of microcrystals was performed using Opentrons OT-2 pipetting robot. The protocol was written using their python API.

*Synthesis for Time-dependent Photoluminescence Measurements*

To study the time-dependent changes in photoluminescence as the microcrystals are exposed to ambient conditions, we synthesized 15 different binary systems. For each system, the exact volumes shown in **Table S2** were deposited into each well. Once each of the precursors was deposited in their respective wells, 250 µL of the chosen antisolvent (chloroform or toluene) was added. We observed an immediate change in color, indicating the formation of microcrystals. As shown in **Figure S1**, there is an apparent difference in the formation, based on the color, of the microcrystals depending on the selection of antisolvent.

|   | 1 | 2 | 3 | 4 | 5 | 6 | 7 | 8 | 9 | 10 | 11 | 12 |
|---|---|---|---|---|---|---|---|---|---|---|---|---|
| **A** | 50 | 49 | 48 | 47.5 | 47 | 46.5 | 46 | 45.5 | 45 | 44.5 | 44 | 43.5 |
|   | 0 | 1 | 2 | 2.5 | 3 | 3.5 | 4 | 4.5 | 5 | 5.5 | 6 | 6.5 |
| **B** | 43 | 42.5 | 42 | 41.5 | 41 | 40.5 | 40 | 39.5 | 39 | 38.5 | 38 | 37.5 |
|   | 7 | 7.5 | 8 | 8.5 | 9 | 9.5 | 10 | 10.5 | 11 | 11.5 | 12 | 12.5 |
| **C** | 37 | 36.5 | 36 | 35.5 | 35 | 34.5 | 34 | 33.5 | 33 | 32.5 | 32 | 31.5 |
|   | 13 | 13.5 | 14 | 14.5 | 15 | 15.5 | 16 | 16.5 | 17 | 17.5 | 18 | 18.5 |
| **D** | 31 | 30.5 | 30 | 29.5 | 29 | 28.5 | 28 | 27.5 | 27 | 26.5 | 26 | 25.5 |
|   | 19 | 19.5 | 20 | 20.5 | 21 | 21.5 | 22 | 22.5 | 23 | 23.5 | 24 | 24.5 |
| **E** | 25 | 24.5 | 24 | 23.5 | 23 | 22.5 | 22 | 21.5 | 21 | 20.5 | 20 | 19.5 |
|   | 25 | 25.5 | 26 | 26.5 | 27 | 27.5 | 28 | 28.5 | 29 | 29.5 | 30 | 30.5 |
| **F** | 19 | 18.5 | 18 | 17.5 | 17 | 16.5 | 16 | 15.5 | 15 | 14.5 | 14 | 13.5 |
|   | 31 | 31.5 | 32 | 32.5 | 33 | 33.5 | 34 | 34.5 | 35 | 35.5 | 36 | 36.5 |
| **G** | 13 | 12.5 | 12 | 11.5 | 11 | 10.5 | 10 | 9.5 | 9 | 8.5 | 8 | 7.5 |
|   | 37 | 37.5 | 38 | 38.5 | 39 | 39.5 | 40 | 40.5 | 41 | 41.5 | 42 | 42.5 |
| **H** | 7 | 6.5 | 6 | 5.5 | 5 | 4.5 | 4 | 3.5 | 3 | 2 | 1 | 0 |
|   | 43 | 43.5 | 44 | 44.5 | 45 | 45.5 | 46 | 46.5 | 47 | 48 | 49 | 50 |

**Table S2.** Exact volume of each precursor deposited into each well. Precursor 1 is shown in black and precursor 2 is shown in red.

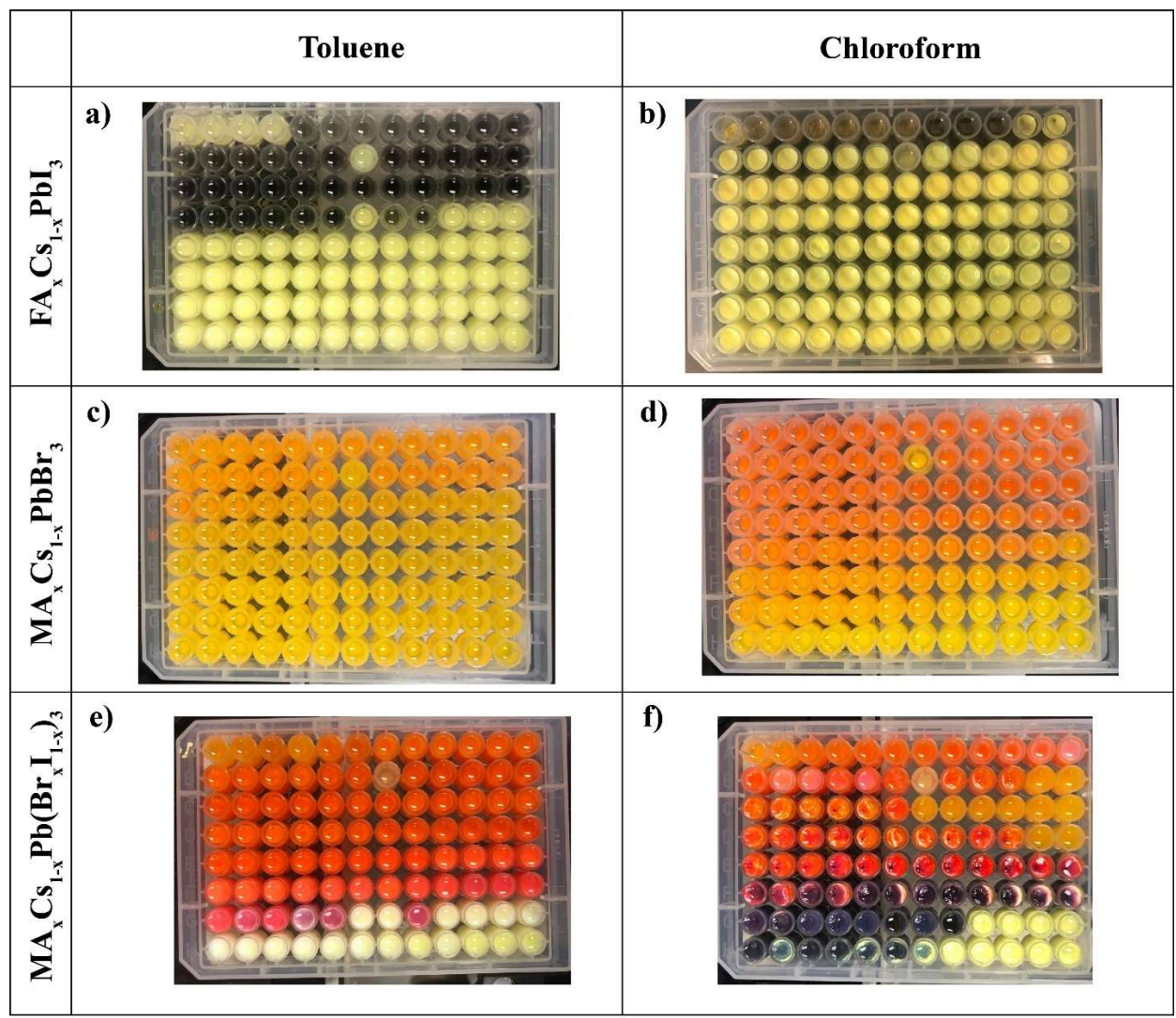

**Figure S1.** Photographs of the microplates when **a)**, **c)**, and **e)** toluene and **b)**, **d)**, and **f)** chloroform is used as the antisolvent for $FA_xCs_{1-x}PbI_3$, $MA_xCs_{1-x}PbBr_3$, and $MA_xCs_{1-x}Pb(Br_xI_{1-x})_3$ systems, respectively.

*Photoluminescence Measurements*

BioTek Cytation Hybrid Multi-Mode Reader is used for photoluminescence spectroscopy. The data collection is designed through Gen 5$^{TM}$ software with capability of data processing for most complex arrays. Photoluminescence measurements were performed with an excitation wavelength of 450 nm and measured over the range of 500 to 850 nm with steps of 5 nm. It was measured from the bottom, approximately 7 mm below the well plate. All measurements were performed in a sweep mode. Photoluminescence measurements were performed every 5 minutes for approximately 6 hours.

**Supplementary Figures**

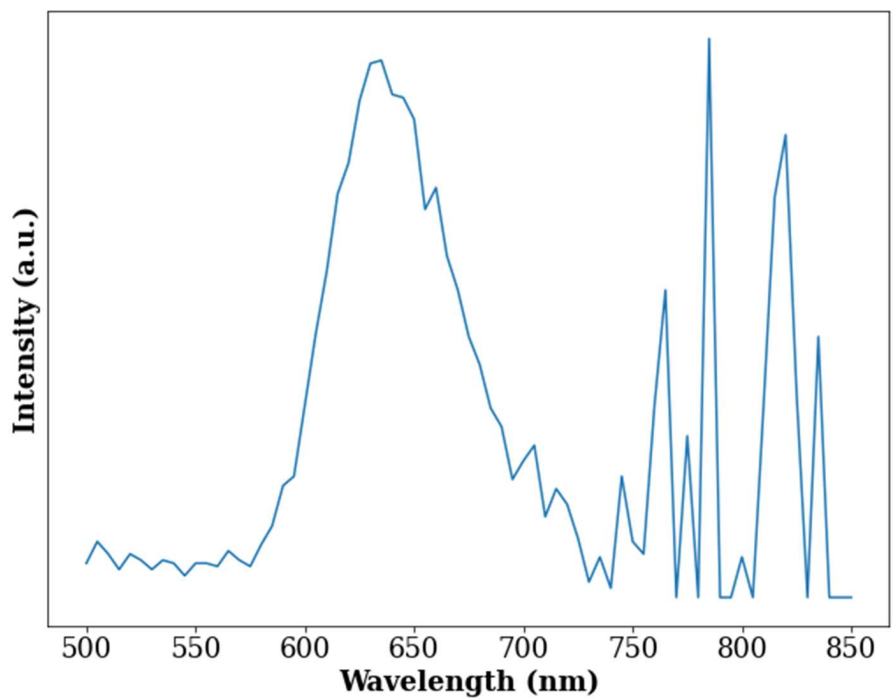

**Figure S2: PL spectrum of FAPbI$_3$.** Observed a broad peak around 650 nm, indicating the formation of the photoinactive, yellow phase (δ-FAPbI$_3$).

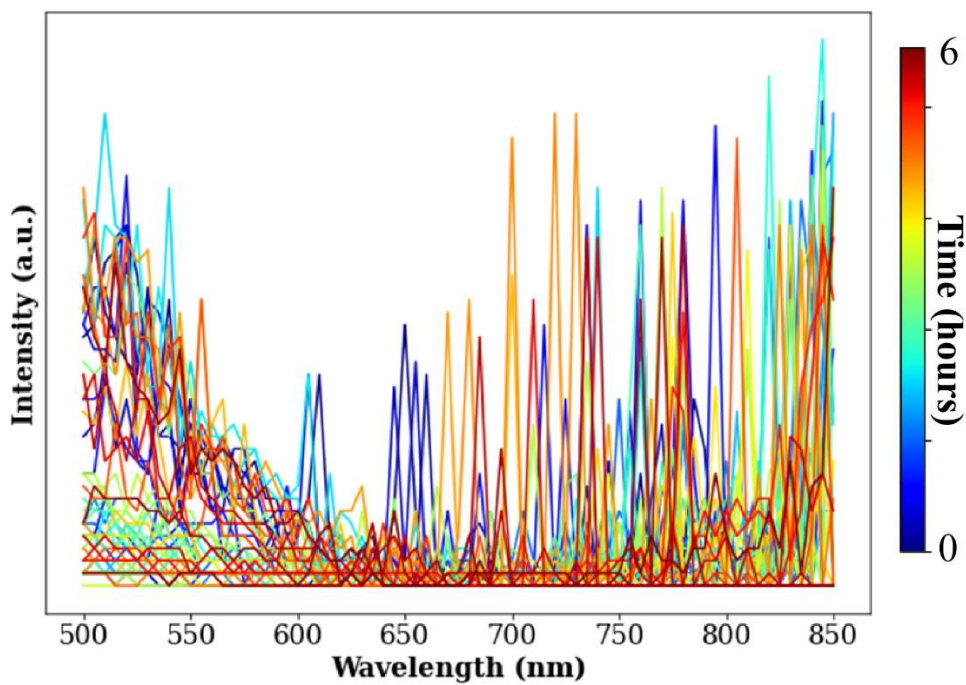

**Figure S3: PL spectrum of MA$_{0.20}$Cs$_{0.80}$Pb(I$_{0.20}$Br$_{0.80}$)$_3$.** Absence of a defined peak, indicating the rapid degradation of CsPbBr$_3$-rich compositions when toluene is used as the antisolvent.

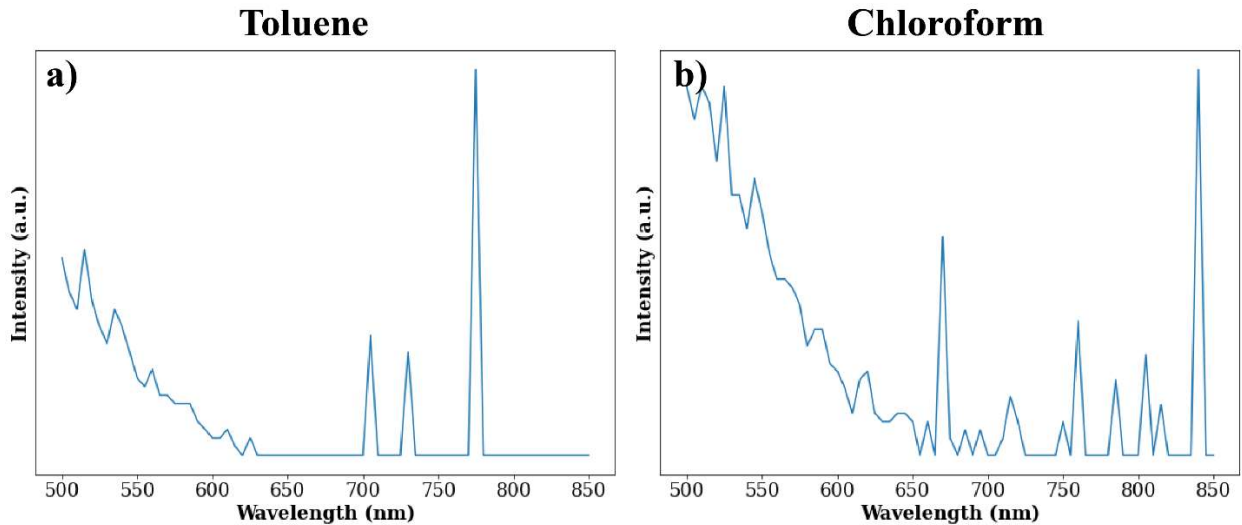

**Figure S4: PL spectra of CsPbI$_3$ when a) toluene and b) chloroform is used as the antisolvent.** Notably, we observe a broad peak at lower wavelengths, indicating the formation of the wide-band gap, polymorph δ-CsPbI$_3$.

*Individual Binary Systems*

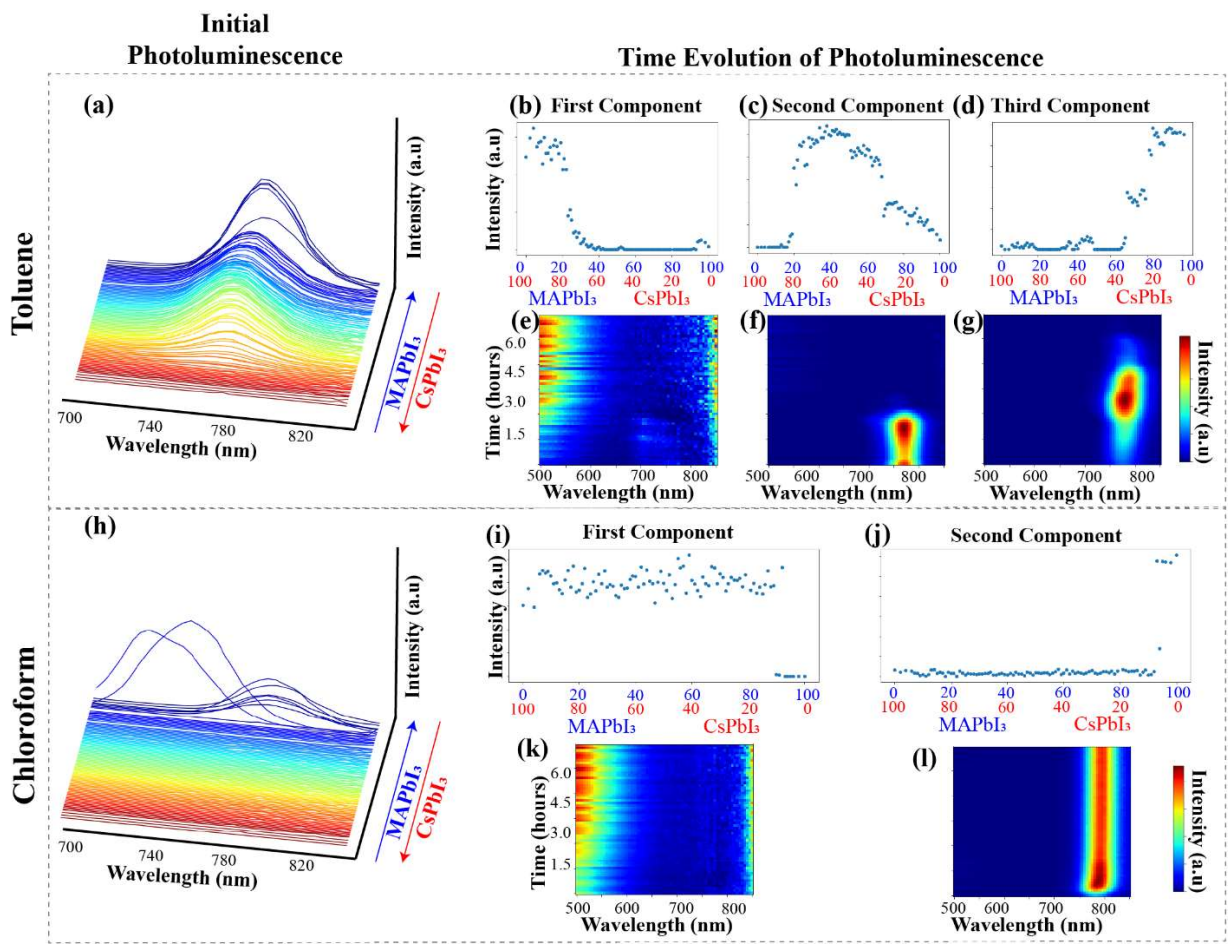

**Figure S5: Double cation lead iodide system, MA$_x$Cs$_{1-x}$PbI$_3$.** Initial PL behavior when **a)** toluene and **h)** chloroform is used as the antisolvent. When toluene is used as the antisolvent, **b)**, **c)**, and **d)** are the loading maps for CsPbI$_3$-rich compositions, solid solutions of CsPbI$_3$ and MAPbI$_3$, and MAPbI$_3$-rich compositions, respectively. **e)**, **f)**, and **g)** Characteristic PL behavior for CsPbI$_3$-rich compositions, solid solutions of CsPbI$_3$ and MAPbI$_3$, and MAPbI$_3$-rich compositions, respectively. When chloroform is used as the antisolvent, **i)** and **j)** and **k)** and **l)** are the loading maps and characteristic PL behavior for CsPbI$_3$-rich compositions, solid solutions of CsPbI$_3$ and MAPbI$_3$, and MAPbI$_3$-rich compositions, respectively.

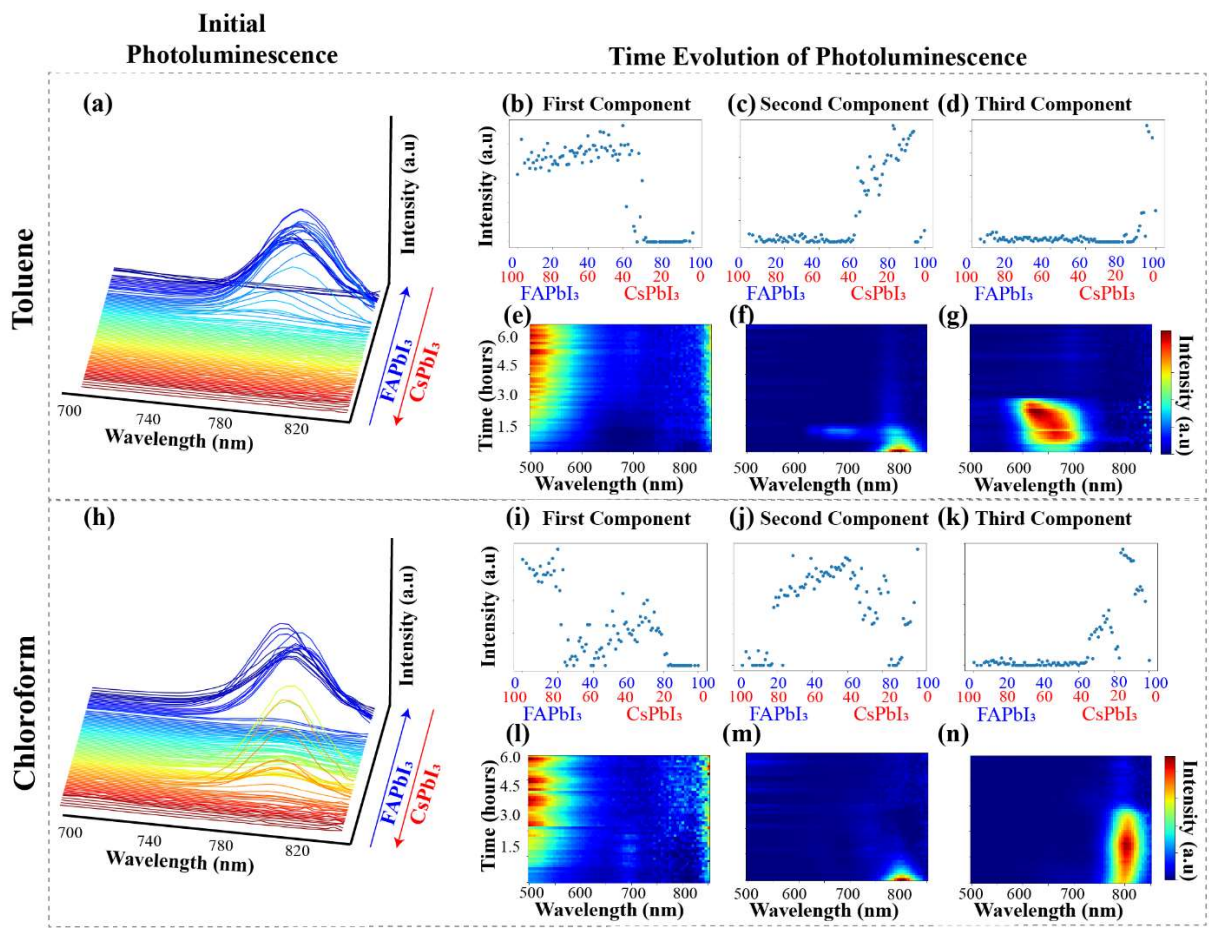

**Figure S6: Double cation lead iodide system, $FA_xCs_{1-x}PbI_3$.** Initial PL behavior when **a)** toluene and **h)** chloroform is used as the antisolvent. When toluene is used as the antisolvent, **b)**, **c)**, and **d)** are the loading maps for $CsPbI_3$-rich compositions, solid solutions of $CsPbI_3$ and $FAPbI_3$, and $FAPbI_3$-rich compositions, respectively. **e)**, **f)**, and **g)** Characteristic PL behavior for $CsPbI_3$-rich compositions, solid solutions of $CsPbI_3$ and $FAPbI_3$, and $FAPbI_3$-rich compositions, respectively. When chloroform is used as the antisolvent, **i)**, **j)** and **k)** and **l)**, **m)**, and **n)** are the loading maps and characteristic PL behavior for $CsPbI_3$-rich compositions, solid solutions of $CsPbI_3$ and $FAPbI_3$, and $FAPbI_3$-rich compositions, respectively.

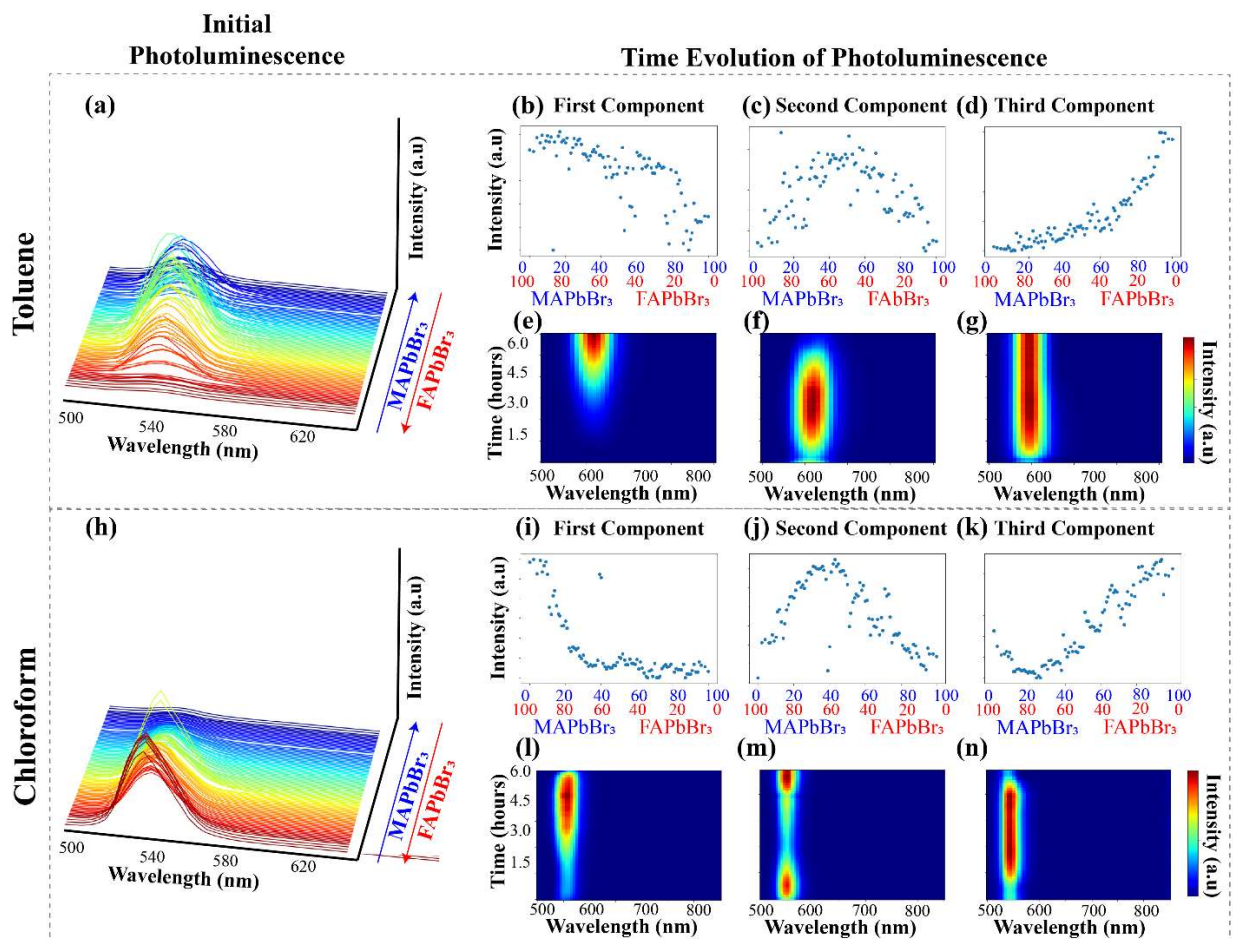

**Figure S7: Double cation lead bromide system, $MA_xFA_{1-x}PbBr_3$.** Initial PL behavior when **a)** toluene and **h)** chloroform is used as the antisolvent. When toluene is used as the antisolvent, **b)**, **c)**, and **d)** are the loading maps for $FAPbBr_3$-rich compositions, solid solutions of $FAPbBr_3$ and $MAPbBr_3$, and $MAPbBr_3$-rich compositions, respectively. **e)**, **f)**, and **g)** Characteristic PL behavior for $FAPbBr_3$-rich compositions, solid solutions of $FAPbBr_3$ and $MAPbBr_3$, and $MAPbBr_3$-rich compositions, respectively. When chloroform is used as the antisolvent, **i)**, **j)** and **k)** and **l)**, **m)**, and **n)** are the loading maps and characteristic PL behavior for $FAPbBr_3$-rich compositions, solid solutions of $FAPbBr_3$ and $MAPbBr_3$, and $MAPbBr_3$-rich compositions, respectively.

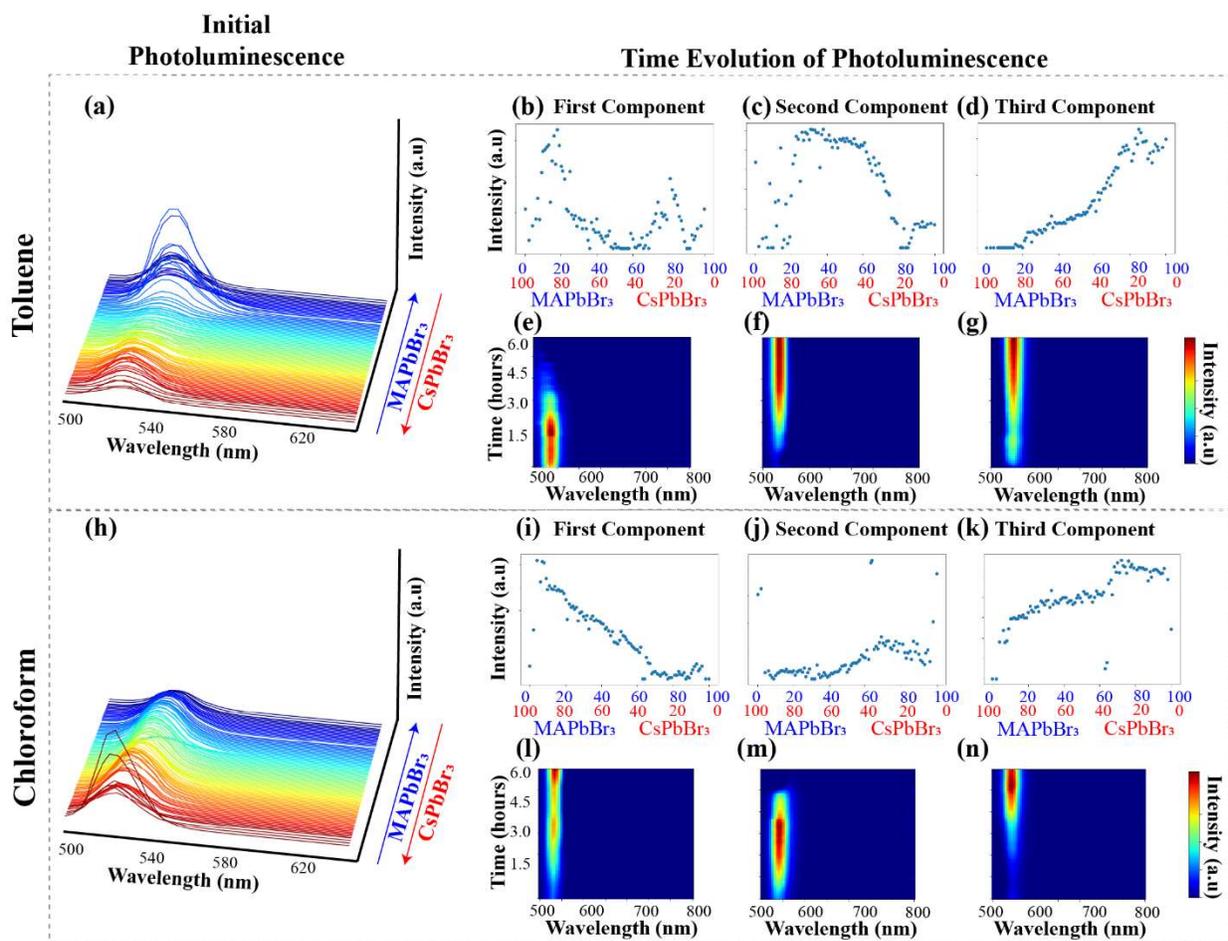

**Figure S8: Double cation lead bromide system, $MA_xCs_{1-x}PbBr_3$.** Initial PL behavior when **a)** toluene and **h)** chloroform is used as the antisolvent. When toluene is used as the antisolvent, **b)**, **c)**, and **d)** are the loading maps for $CsPbBr_3$-rich compositions, solid solutions of $CsPbBr_3$ and $MAPbBr_3$, and $MAPbBr_3$-rich compositions, respectively. **e)**, **f)**, and **g)** Characteristic PL behavior for $CsPbBr_3$-rich compositions, solid solutions of $CsPbBr_3$ and $MAPbBr_3$, and $MAPbBr_3$-rich compositions, respectively. When chloroform is used as the antisolvent, **i)**, **j)** and **k)** and **l)**, **m)**, and **n)** are the loading maps and characteristic PL behavior for $CsPbBr_3$-rich compositions, solid solutions of $CsPbBr_3$ and $MAPbBr_3$, and $MAPbBr_3$-rich compositions, respectively.

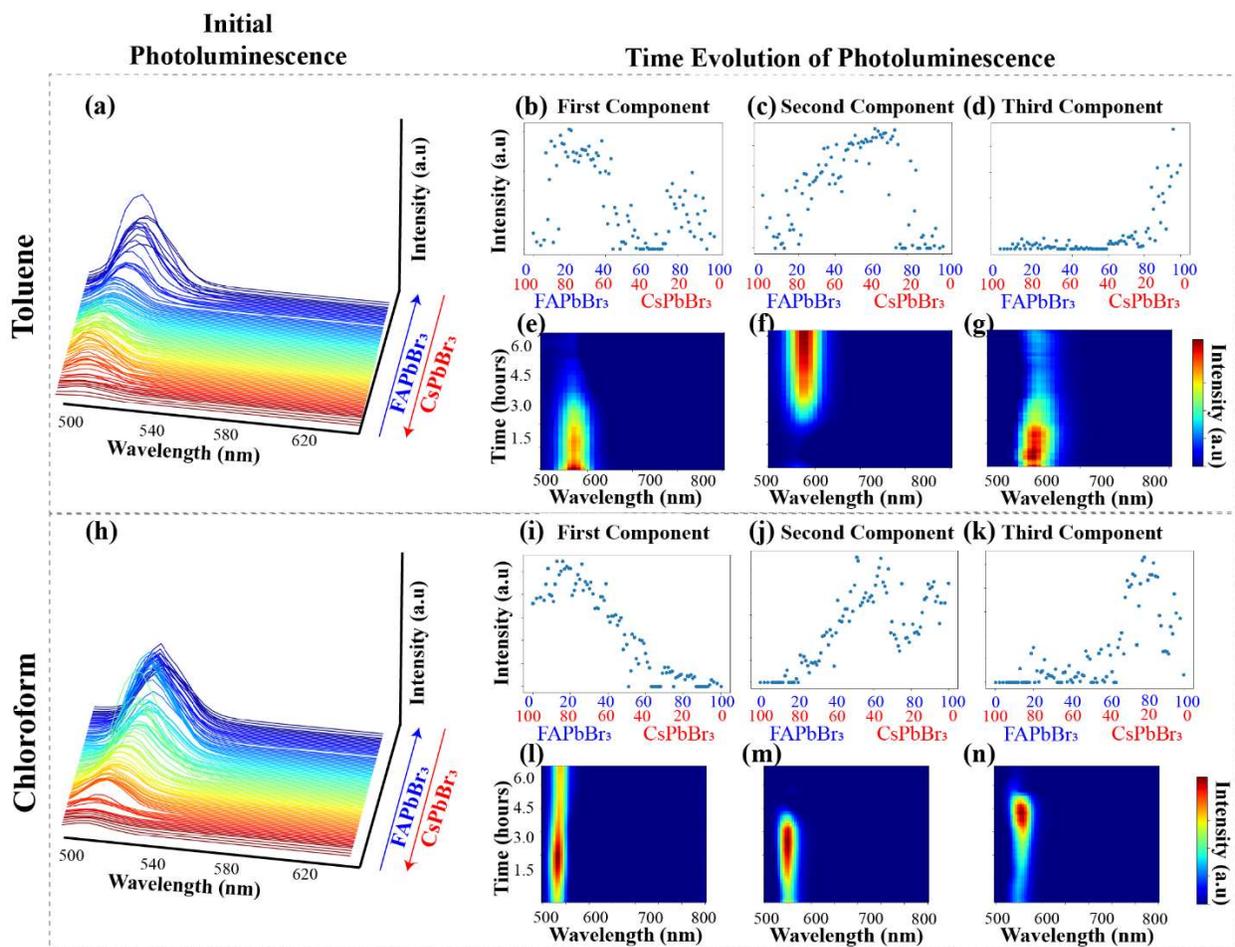

**Figure S9: Double cation lead bromide system, $FA_xCs_{1-x}PbBr_3$.** Initial PL behavior when **a)** toluene and **h)** chloroform is used as the antisolvent. When toluene is used as the antisolvent, **b)**, **c)**, and **d)** are the loading maps for $CsPbBr_3$-rich compositions, solid solutions of $CsPbBr_3$ and $FAPbBr_3$, and $FAPbBr_3$-rich compositions, respectively. **e)**, **f)**, and **g)** Characteristic PL behavior for $CsPbBr_3$-rich compositions, solid solutions of $CsPbBr_3$ and $FAPbBr_3$, and $FAPbBr_3$-rich compositions, respectively. When chloroform is used as the antisolvent, **i)**, **j)** and **k)** and **l)**, **m)**, and **n)** are the loading maps and characteristic PL behavior for $CsPbBr_3$-rich compositions, solid solutions of $CsPbBr_3$ and $FAPbBr_3$, and $FAPbBr_3$-rich compositions, respectively.

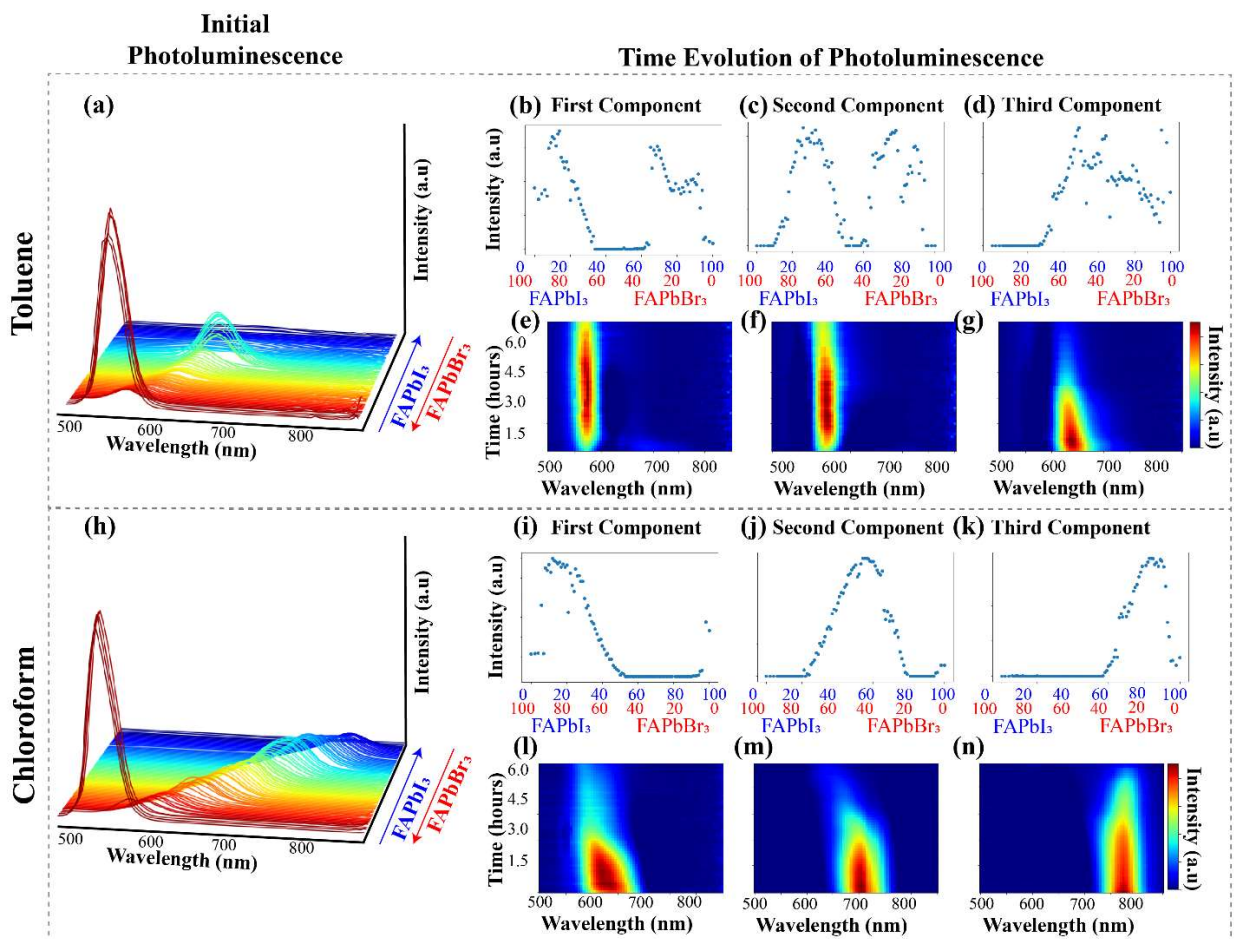

**Figure S10: Formamidinium lead double halide system, FAPb($I_xBr_{1-x}$)$_3$.** Initial PL behavior when **a)** toluene and **h)** chloroform is used as the antisolvent. When toluene is used as the antisolvent, **b)**, **c)**, and **d)** are the loading maps for FAPbBr$_3$-rich compositions, solid solutions of FAPbBr$_3$ and FAPbI$_3$, and FAPbI$_3$-rich compositions, respectively. **e)**, **f)**, and **g)** Characteristic PL behavior for FAPbBr$_3$-rich compositions, solid solutions of FAPbBr$_3$ and FAPbI$_3$, and FAPbI$_3$-rich compositions, respectively. When chloroform is used as the antisolvent, **i)**, **j)** and **k)** and **l)**, **m)**, and **n)** are the loading maps and characteristic PL behavior for FAPbBr$_3$-rich compositions, solid solutions of FAPbBr$_3$ and FAPbI$_3$, and FAPbI$_3$-rich compositions, respectively.

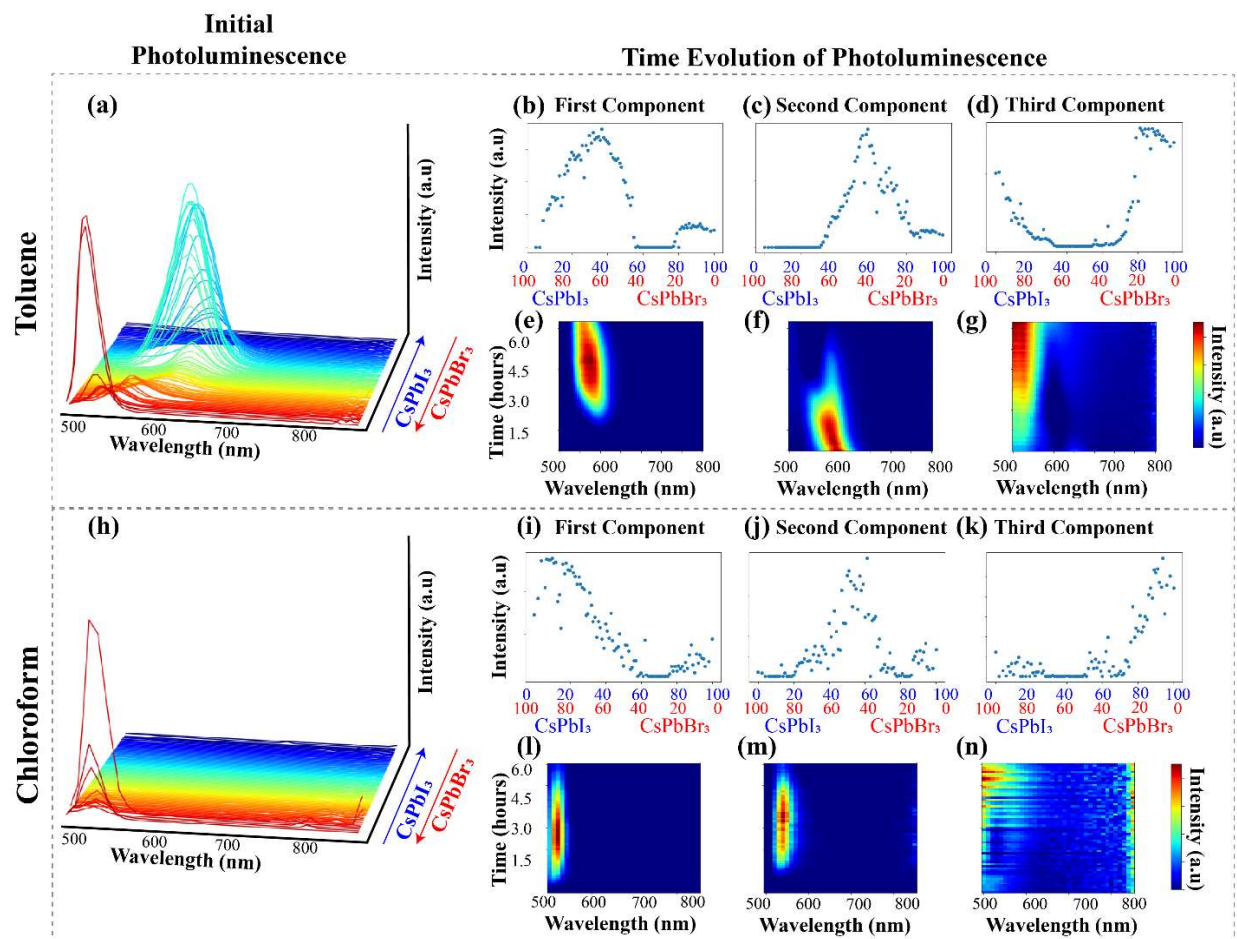

**Figure S11: Cesium lead double halide system, $CsPb(I_xBr_{1-x})_3$.** Initial PL behavior when **a)** toluene and **h)** chloroform is used as the antisolvent. When toluene is used as the antisolvent, **b)**, **c)**, and **d)** are the loading maps for $CsPbBr_3$-rich compositions, solid solutions of $CsPbBr_3$ and $CsPbI_3$, and $CsPbI_3$-rich compositions, respectively. **e)**, **f)**, and **g)** Characteristic PL behavior for $CsPbBr_3$-rich compositions, solid solutions of $CsPbBr_3$ and $CsPbI_3$, and $CsPbI_3$-rich compositions, respectively. When chloroform is used as the antisolvent, **i)**, **j)** and **k)** and **l)**, **m)**, and **n)** are the loading maps and characteristic PL behavior for $CsPbBr_3$-rich compositions, solid solutions of $CsPbBr_3$ and $CsPbI_3$, and $CsPbI_3$-rich compositions, respectively.

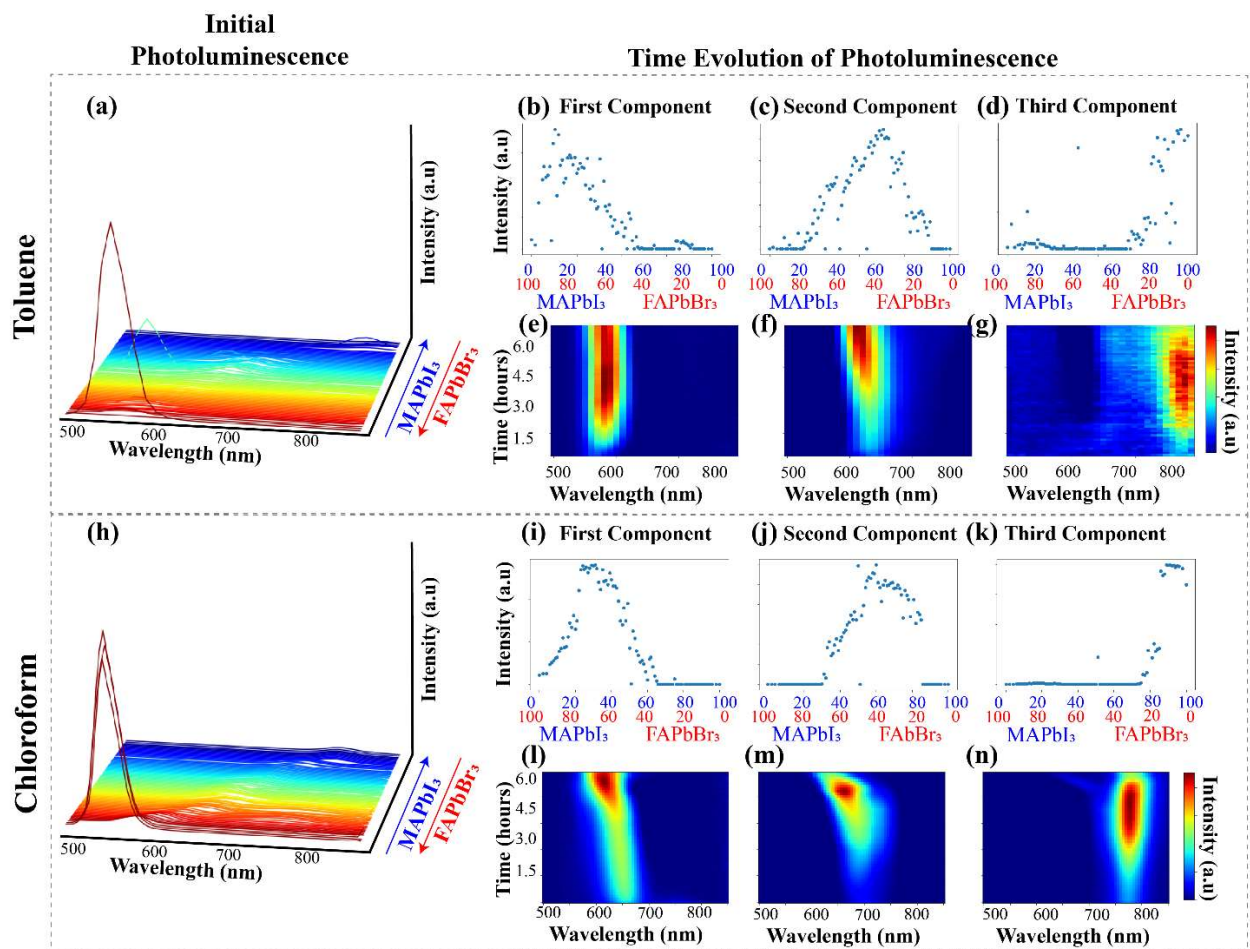

**Figure S12: Double cation lead double halide system, $MA_xFA_{1-x}Pb(I_xBr_{1-x})_3$.** Initial PL behavior when **a)** toluene and **h)** chloroform is used as the antisolvent. When toluene is used as the antisolvent, **b)**, **c)**, and **d)** are the loading maps for FAPbBr₃-rich compositions, solid solutions of FAPbBr₃ and MAPbI₃, and MAPbI₃-rich compositions, respectively. **e)**, **f)**, and **g)** Characteristic PL behavior for FAPbBr₃-rich compositions, solid solutions of FAPbBr₃ and MAPbI₃, and MAPbI₃-rich compositions, respectively. When chloroform is used as the antisolvent, **i)**, **j)** and **k)** and **l)**, **m)**, and **n)** are the loading maps and characteristic PL behavior for FAPbBr₃-rich compositions, solid solutions of FAPbBr₃ and MAPbI₃, and MAPbI₃-rich compositions, respectively.

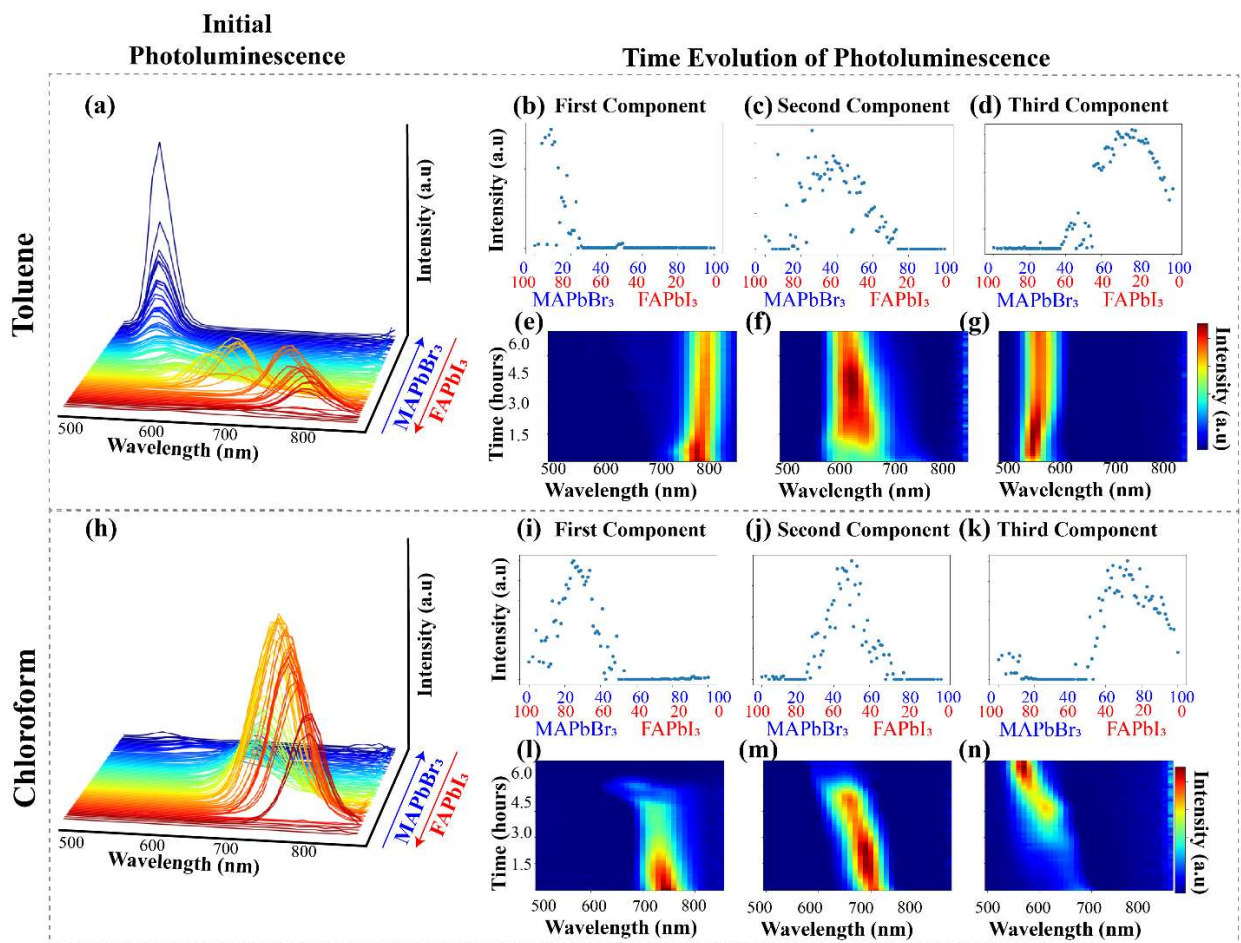

**Figure S13: Double cation lead double halide system, $MA_xFA_{1-x}Pb(Br_xI_{1-x})_3$.** Initial PL behavior when **a)** toluene and **h)** chloroform is used as the antisolvent. When toluene is used as the antisolvent, **b)**, **c)**, and **d)** are the loading maps for FAPbI$_3$-rich compositions, solid solutions of FAPbI$_3$ and MAPbBr$_3$, and MAPbBr$_3$-rich compositions, respectively. **e)**, **f)**, and **g)** Characteristic PL behavior for FAPbI$_3$-rich compositions, solid solutions of FAPbI$_3$ and MAPbBr$_3$, and MAPbBr$_3$-rich compositions, respectively. When chloroform is used as the antisolvent, **i)**, **j)** and **k)** and **l)**, **m)**, and **n)** are the loading maps and characteristic PL behavior for FAPbI$_3$-rich compositions, solid solutions of FAPbI$_3$ and MAPbBr$_3$, and MAPbBr$_3$-rich compositions, respectively.

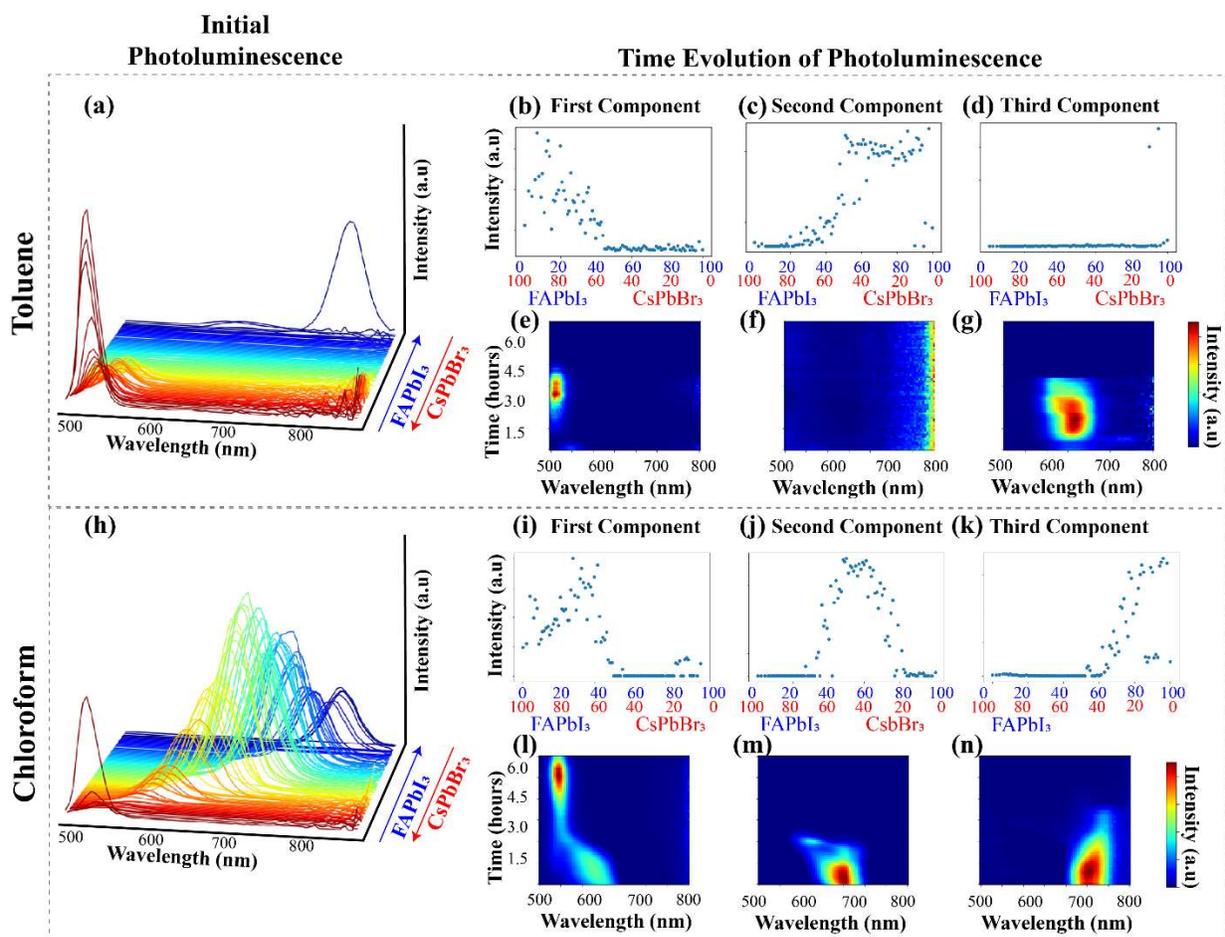

**Figure S14: Double cation lead double halide system, $FA_xCs_{1-x}Pb(I_xBr_{1-x})_3$.** Initial PL behavior when **a)** toluene and **h)** chloroform is used as the antisolvent. When toluene is used as the antisolvent, **b)**, **c)**, and **d)** are the loading maps for $CsPbBr_3$-rich compositions, solid solutions of $CsPbBr_3$ and $FAPbI_3$, and $FAPbI_3$-rich compositions, respectively. **e)**, **f)**, and **g)** Characteristic PL behavior for $CsPbBr_3$-rich compositions, solid solutions of $CsPbBr_3$ and $FAPbI_3$, and $FAPbI_3$-rich compositions, respectively. When chloroform is used as the antisolvent, **i)**, **j)** and **k)** and **l)**, **m)**, and **n)** are the loading maps and characteristic PL behavior for $CsPbBr_3$-rich compositions, solid solutions of $CsPbBr_3$ and $FAPbI_3$, and $FAPbI_3$-rich compositions, respectively.

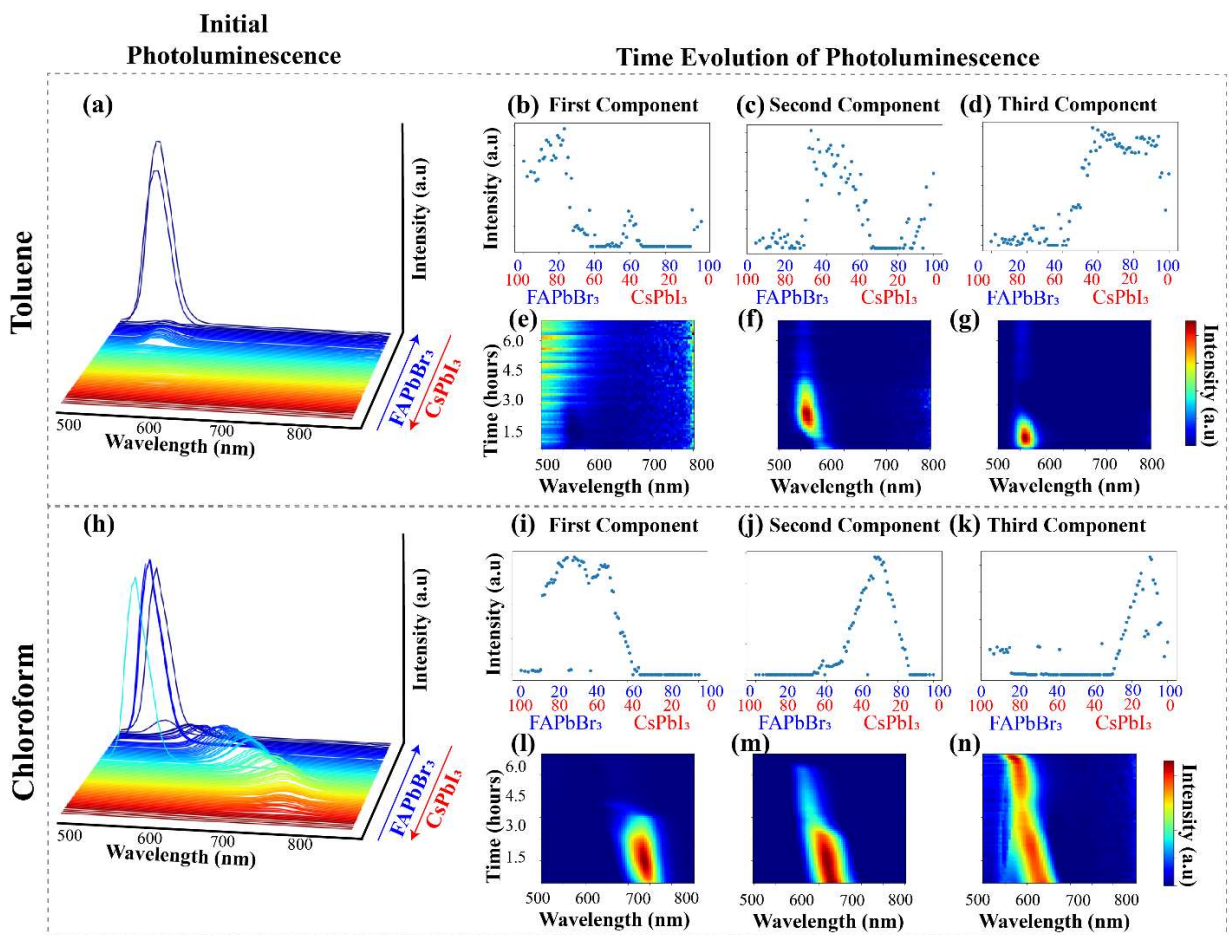

**Figure S15: Double cation lead double halide system, $FA_xCs_{1-x}Pb(Br_xI_{1-x})_3$.** Initial PL behavior when **a)** toluene and **h)** chloroform is used as the antisolvent. When toluene is used as the antisolvent, **b)**, **c)**, and **d)** are the loading maps for $CsPbI_3$-rich compositions, solid solutions of $CsPbI_3$ and $FAPbBr_3$, and $FAPbBr_3$-rich compositions, respectively. **e)**, **f)**, and **g)** Characteristic PL behavior for $CsPbI_3$-rich compositions, solid solutions of $CsPbI_3$ and $FAPbBr_3$, and $FAPbBr_3$-rich compositions, respectively. When chloroform is used as the antisolvent, **i)**, **j)** and **k)** and **l)**, **m)**, and **n)** are the loading maps and characteristic PL behavior for $CsPbI_3$-rich compositions, solid solutions of $CsPbI_3$ and $FAPbBr_3$, and $FAPbBr_3$-rich compositions, respectively.